\documentclass[oldversion]{aa} 

\usepackage{graphicx}
\usepackage{txfonts}
\usepackage{natbib}
\bibpunct{(}{)}{;}{a}{}{,}

\usepackage{ifthen}
\usepackage[figuresright]{rotating}
\usepackage{longtable}
\usepackage{supertabular}

\begin{document}
   \title{A search for diffuse bands in the circumstellar\\ envelopes of
           post-AGB stars\thanks{Based on observations collected at the
  European Southern Observatory (Chile) and at the Spanish Observatorio del
  Roque de los Muchachos of the Instituto de Astrof\'\i sica de Canarias.}}
   \author{R. Luna\inst{1}
        \and
           N.L.J. Cox \inst{2}
          \and
          M.A. Satorre \inst{1}
          \and
          D. A. Garc\'ia Hern\'andez\inst{3}
          \and
          O. Su\'arez \inst{4}
          \and
          P. Garc\'ia Lario \inst{2}
          }

   \offprints{R. Luna, \email{ralunam@fis.upv.es}}

   \institute{Laboratorio de Astrof\'\i sica Experimental.
     Escuela Polit\'ecnica Superior de Alcoy,
     Universidad Polit\'ecnica de Valencia,
     Plaza de Ferr\'andiz y Carbonell,
     E-03801 Alcoy, Alicante, Spain
     \and
     Herschel Science Centre.  Research and Scientific Support Department of ESA,
     European Space Astronomy Centre, P.O.\,Box 78, 28691,
     Villanueva de la Ca\~nada, Madrid, Spain
     \and
     W. J. McDonald Observatory. The University of Texas at Austin.
     1 University Station, C1400. Austin, TX 78712-0259, USA
     \and
     LUAN, Universit\'e de Nice Sophia Antipolis, Parc Valrose, 
     06108 Nice cedex 2, France}

   \date{Received 27 March 2006; accepted 6 November 2007}

\abstract{
In this work we present the results of a systematic search for diffuse bands
(DBs, hereafter) in the circumstellar envelopes of a carefully selected sample
of post-AGB stars. We concentrated on the analysis of 9 of the DBs most commonly
found in the interstellar medium. The strength of these features is determined
using high resolution optical spectroscopy and the results obtained are
compared with literature data on field stars affected only by interstellar
reddening. Based on the weak features observed in the subsample of post-AGB
stars dominated by circumstellar reddening we conclude that the carrier(s) of
these DBs must not be present in the circumstellar environment of these
sources, or at least not under the excitation conditions in which DBs are
formed. The conclusion is applicable to all the post-AGB stars studied,
irrespective of the dominant chemistry or the spectral type of the star
considered. A detailed radial velocity analysis of the features observed in
individual sources confirms this result, as the Doppler shifts measured are
found to be consistent with an interstellar origin.
}

\titlerunning{A search for diffuse bands in post-AGB stars}

   \maketitle
%

\section{Introduction}
The diffuse interstellar bands (DIBs) are absorption features, 
showing a broad range of widths and strengths, which appear over-imposed
on the spectra of bright stars whose lines of sight probe (extra)galactic
diffuse to dense interstellar clouds. 
Currently, more than 200 DIBs have been
identified and catalogued in the spectral range from 3600 to 10200~\AA\
\citep{Jenniskens94b,Cox05}, the most studied ones being those found at 4430,
5780, 5797 and 6284~\AA. Since their discovery \citep{Heger22}, they have been
associated to the interstellar medium (ISM), because their strengths show a
positive relationship with the observed extinction \citep{Merrill36} as well as
to the neutral sodium column density \citep{Herbig93}. Many carriers have been
proposed, however, no unambiguous identification has yet been made and it is
debated whether they arise from the dust or the gas component of the ISM (see
reviews by \citealt{Herbig95} and \citealt{Sarre06}). There is increasing
observational evidence that the DIB carriers constitute a set of carbonaceous
gas phase molecules as evidenced from substructures resembling rotational
contours in some bands (\citealt{Sarre95}, \citealt{Ehrenfreund96}). In
particular, photo-UV-resistant organic molecules, such as carbon chains
\citep{Douglas77}, PAHs \citep{Salama99,Allamandola99}, fullerenes
\citep{Foing94,Iglesias-Groth07} and / or buckyonions \citep{Iglesias-Groth04}
are promising candidates. The local interstellar environmental conditions
set the balance of local formation and destruction of the carriers as well as
their level of ionization and hydrogenation. The interstellar radiation field
is one of the most important factors in this \citep{Ruiterkamp05}.

There is a possible link between the DIB carriers and the carriers of the
unidentified (aromatic) infrared bands (UIBs), the so-called PAH-DIB hypothesis
(\citealt{Crawford85, LegerDHendecours85, VanderZwet-Allamandola85}).

Thus, although PAHs are thought to reside and to be processed (ionisation,
dehydrogenation, destruction) in the diffuse ISM, this does not exclude the
scenario that these molecules (or their parent species) are produced elsewhere.
Since circumstellar shells are sources of replenishment of the ISM, it has been
argued that DIBs (and/or parent structures) may have a circumstellar origin,
either in dense stellar winds or circumstellar shells, thus somehow
contravening the name they were initially given. The suspected connection
between DIB carriers and some carbon-rich compounds can be investigated
attending to the usually known chemistry and physical properties of these
circumstellar shells.

Observationally, the detection of diffuse bands (DBs, hereafter) 
around evolved
stars is hampered by the fact that most of them are mass-losing stars, usually
strongly variable, and surrounded by very cool extended atmospheres where
molecules are the dominant source of opacity. These stars are very difficult to
model and DBs are hardly detected (in absorption) against the forest of features attributed to
molecular transitions which appear over-imposed on the stellar continuum. This
has hampered the systematic search for DBs in evolved stars in the past.
Furthermore, if detected, it is difficult to determine whether the DBs are
originating from the interstellar or the circumstellar environment, or even
both.

In face of these difficulties \citet{SnowWallerstein73} and \citet{Snow73}
searched for circumstellar diffuse bands (at 4430, 5780, 5797 and 6614~\AA) in
26 stars with suspected circumstellar dust shells / envelopes but found no
evidence for their presence. Several other authors have since searched for and
commented on the presence of diffuse bands and their interstellar or
circumstellar nature, in spectra observed towards planetary nebulae
(\mbox{\object{NGC 6210}}, \mbox{\object{NGC 7027}}, \mbox{\object{IC 351}} and
\mbox{\object{AFGL 2688}} by \citealt{PritchetGrillmair84}; \mbox{\object{IRAS
21282+5050}} by \citealt{CohenJones87}; \mbox{\object{NGC 7027}} and
\mbox{\object{IRAS 21282+5050}} by \citealt{LeBertreLequeux92}), a post-AGB
star (\mbox{\object{HR 4049}}; \citealt{WatersEtal89}), and a carbon star
(\mbox{\object{IRAS 07270-1921}} or \mbox{CGCS 1732};
\citealt{LeBertre90}). \citet{Lebertre93} studied a new sample consisting of
carbon-, oxygen- or nitrogen-rich mass-losing sources, such as (pre)planetary
nebulae (\mbox{\object{BD+30 3639}}, \mbox{\object{CPD-56 8032}},
\mbox{Hen 104}), a carbon rich RV Tauri star (\mbox{\object{AC Her}}), wolf-rayet
stars (\mbox{\object{WR137}}, \mbox{\object{WR140}}) and post-AGB stars
(\mbox{\object{HR 4049}}, \mbox{\object{HD213985}}), revisiting several sources
studied in the past (\emph{e.g.} \mbox{CS 776}, \mbox{\object{NCG 7027}},
\mbox{\object{HR 4049}}, \mbox{\object{IRAS 21282+5050}}). These
authors did not find any evidence for circumstellar DBs and 
refuted previous
claims and thus concluded that DBs are depleted in circumstellar environments.
Notably, these authors did not detect any bands in the spectra of sources
showing strong PAH emission (UIB) at mid-infrared wavelengths. This suggested
that carrier molecules, if present, in circumstellar envelopes are in a
different state of ionization / hydrogenation than in the ISM. Strong DBs were
detected toward carbon-rich sources that do not show PAH emission, as well as
toward most of their oxygen-rich and nitrogen-rich sources in the sample,
although in all cases the observed DBs could be attributed to the interstellar
material in the lines of sight. For unexplained reasons enhanced DBs were
detected toward WN stars and LBVs.
Exceptionally, narrow emission features possibly related to DIBs
as well have been observed toward the \object{Red Rectangle} 
\citep{Scarrott92}, although their identification and nature remains controversial.

A largely unexplored alternative exists. This concerns the so-called post-AGB
stars that are in a short-lived transition phase between the Asymptotic Giant
Branch (AGB) and the Planetary Nebula (PN) stage, evolving very rapidly in the
H-R diagram while they are still surrounded by the remnant of the AGB
circumstellar shell. Post-AGB stars show all possible spectral types from M to
B in what probably represents an evolutionary sequence of increasing effective
temperature in their way to become PNe \citep{Garcia-Lario97b}. This means that
in these stars we should easily be able to detect DBs formed in the remnant AGB
shell over-imposed on the intermediate or early-type spectrum of the central
star without the confusion originating from the presence of molecular bands in
AGB stars. Interestingly, while many of these DBs are common to those observed
in the ISM, the relative ratios are sometimes found to be very different
\citep{Garcia-Lario99}. These circumstellar DBs could form and survive for some
time under conditions which might be substantially different to those found in
the ISM in terms of density, UV radiation field, etc. and could hold the key to
understand and solve this long-standing problem. Another advantage is the fact
that the chemical composition of the gas and dust in these shells can easily be
determined from observations in the optical, infrared, mm / sub-mm or radio
wavelengths. In addition, post-AGB stars are located in many cases at relative
high galactic latitudes, and are as such affected only by little interstellar
reddening. This facilitates the attribution of a circumstellar origin to the
features observed.

The potential formation of DBs around post-AGB stars has, however, been
explored so far only occasionally for a limited number of sources.
Nevertheless, the presence of strong DBs has been reported in the optical
spectra of a few post-AGB stars
(\citealt{Lebertre93,Garcia-Lario99,Zacs99,Zacs01,Klochkova99,Klochkova00,Kendall02})
and some carbon rich (barium) stars \citep{Zacs03}. In several cases tentative
claims have been put forward of DBs detected at radial velocities coinciding
with the photospheric absorption lines or shell/envelope expansion velocity
(\mbox{\object{IRAS 04296+3429}} by \citealt{Klochkova99}; and \object{HD
179821} by \citealt{Zacs99}). All other studies listed above gave
non-conclusive results.

Another method recently employed utilised nearby background stars to probe the
circumstellar environments of the carbon star \mbox{\object{IRC+10216}}
\citep{Kendall02} and the Helix PN \citep{MauronKendall04}. Again, no
circumstellar DBs were detected, confirming the lack of DIB carriers in these
environments.


\begin{table*}[ht!]
\caption{Observation log.}\centering
\label{tb:observations}
\resizebox{\textwidth}{!}{
\begin{tabular}{ l c c c c c c c }\hline\hline
Name                  &      Other Names          & Observing Date&   Observatory &   Telescope   & Instrument& Spectral range (\AA) & Resolution\\
\hline
\object{IRAS 01005+7910}  &               &   12.09.03    &   La Palma    &   TNG (3.58 m)&   SARG    &   4960$-$10110    &   50000   \\
\object{IRAS Z02229+6208} &               &   12.09.03    &   La Palma    &   TNG (3.58 m)&   SARG    &   4960$-$10110    &   50000   \\
\object{IRAS 04296+3429}  &               &   22.08.94    &   La Palma    &   WHT (4.2 m) &   UES     &   5570$-$10200    &   55000   \\
\object{IRAS 05113+1347}  &               &   23.02.94    &   La Palma    &   WHT (4.2 m) &   UES     &   4480$-$10200    &   55000   \\
\object{IRAS 05251-1244}  & \object{IC 418}       &   12.09.03    &   La Palma    &   TNG (3.58 m)&   SARG    &   4960$-$10110    &   50000   \\
\object{IRAS 05341+0852}  &               &   30.09.98    &   ESO-La Silla&   NTT (3.58 m)&   EMMI    &   5570$-$10040    &   65000   \\
\object{IRAS 06530-0213}  &               &   12.01.01    &   ESO-Paranal &   VLT-U2 (8 m)&   UVES    &   4790$-$6810     &   100000  \\
\object{IRAS 07134+1005}  & \object{HD 56126}         &   20.12.93    &   La Palma    &   WHT (4.2 m) &   UES     &   3980$-$10400    &   55000   \\
\object{IRAS 08005-2356}  &               &   02.03.94    &   La Palma    &   WHT (4.2 m) &   UES     &   3690$-$11050    &   55000   \\
\object{IRAS 08143-4406}  &               &   16.01.01    &   ESO-Paranal &   VLT-U2 (8 m)&   UVES    &   3750$-$10520    &   100000  \\
\object{IRAS 08544-4431}  &               &   26.01.99    &   ESO-La Silla&   ESO-1.52    &   FEROS   &   3700$-$8860     &   50000   \\
\object{IRAS 12175-5338}  & \object{SAO 239853}       &   24.03.00    &   ESO-La Silla&   ESO-1.52    &   FEROS   &   3730$-$8850     &   50000   \\
\object{IRAS 16594-4656}  &               &   20.05.00    &   ESO-Paranal &   VLT-U2 (8 m)&   UVES    &   3750$-$10520    &   100000  \\
\object{IRAS 17086-2403}  &               &   14.07.01    &   La Palma    &   WHT (4.2 m) &   UES     &   4300$-$9000     &   50000   \\
\object{IRAS 17097-3210}  & \object{HD 155448}        &   07.07.97    &   ESO-La Silla&   NTT (3.58 m)&   EMMI    &   5800$-$10430    &   60000   \\
\object{IRAS 17150-3224}  & \object{RAFGL 6815}       &   10.06.00    &   ESO-Paranal &   VLT-U2 (8 m)&   UVES    &   3750$-$10520    &   100000  \\
\object{IRAS 17245-3951}  &               &   21.05.00    &   ESO-Paranal &   VLT-U2 (8 m)&   UVES    &   3750$-$10520    &   100000  \\
\object{IRAS 17395-0841}  &               &   13.07.01    &   La Palma    &   WHT (4.2 m) &   UES     &   4300$-$9000     &   55000   \\
\object{IRAS 17423-1755}  & \object{Hen 3-1475}       &   13.07.01    &   La Palma    &   WHT (4.2 m) &   UES     &   4300$-$9000     &   55000   \\
\object{IRAS 17436+5003}  & \object{HD 161796}        &   01.09.94    &   La Palma    &   WHT (4.2 m) &   UES     &   3650$-$10200    &   55000   \\
\object{IRAS 18025-3906}  &               &   30.09.98    &   ESO-La Silla&   NTT (3.58 m)&   EMMI    &   3980$-$10430    &   65000   \\
\object{IRAS 18062+2410}  & \object{HD 341617}        &   13.07.01    &   La Palma    &   WHT (4.2 m) &   UES     &   4300$-$9000     &   50000   \\
                          & \object{HD 172324}        &   08.08.95    &   ESO-La Silla&   NTT (3.58 m)&   EMMI    &   3650$-$10240   &   55000   \\
\object{IRAS 19114+0002}  & \object{HD 179821}        &   07.07.97    &   ESO-La Silla&   NTT (3.58 m)&   EMMI    &   5800$-$10430    &   60000   \\
\object{IRAS 19200+3457}  & \object{LS II +34 1}      &   11.09.03    &   La Palma    &   TNG (3.58 m)&   SARG    &   4960$-$10110    &   50000   \\
\object{IRAS 19386+0155}  &               &   30.09.98    &   ESO-La Silla&   NTT (3.58 m)&   EMMI    &   5980$-$8320     &   65000   \\
\object{IRAS 19500-1709}  & \object{HD 187885}        &   07.08.95    &   ESO-La Silla&   NTT (3.58 m)&   EMMI    &   3650$-$10040    &   55000   \\
\object{IRAS 20000+3239}  &               &   23.08.94    &   La Palma    &   WHT (4.2 m) &   UES     &   4440$-$10040    &   50000   \\
\object{IRAS 20462+3416}  & \object{LS II +34 26}     &   17.08.96    &   La Palma    &   WHT (4.2 m) &   UES     &   5300$-$9380     &   55000   \\
\object{IRAS 22023+5249}  & \object{LS III +52 24}    &   14.07.01    &   La Palma    &   WHT (4.2 m) &   UES     &   4300$-$9000     &   55000   \\
\object{IRAS 22223+4327}  & \object{BD+42 4388}       &   24.08.94    &   La Palma    &   WHT (4.2 m) &   UES     &   4440$-$10040    &   50000   \\
\object{IRAS 22272+5435}  & \object{HD 235858}        &   23.08.94    &   La Palma    &   WHT (4.2 m) &   UES     &   4440$-$10220    &   55000   \\
\object{IRAS 23304+6147}  &               &   23.08.94    &   La Palma    &   WHT (4.2 m) &   UES     &   4440$-$10040    &   55000   \\
\hline
\end{tabular}
}
\end{table*}


\begin{figure}[th!]
\begin{center}
\includegraphics[width=\columnwidth]{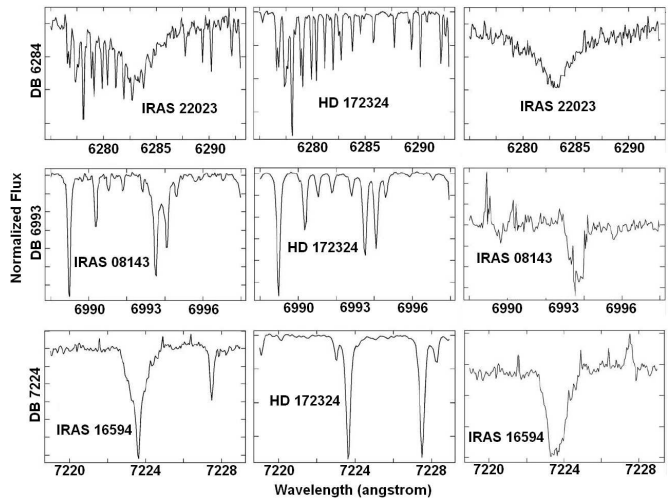}
\caption{
  Removal of telluric lines at 6284, 6993 and 7224~\AA.
  A few examples are shown before (left) and after correction
  (right). In the middle panel we show the stellar spectrum of
  \object{HD 172324} (B9Ib), one of the sample stars, which was
  used as telluric divisor (see text).}
  \label{fig:mezcla}
\end{center}
\end{figure}


\begin{table*}[ht!]
\caption{Main characteristics of the post-AGB stars selected for the analysis} \centering
\label{tb:referencias}
\begin{tabular}{l r@{~$\pm$~}l c c c c c r r }\hline\hline
 IRAS Name    & \multicolumn {2}{c}{$E(B-V)$}    &    Ref.      & Chemistry & Ref.&  Sp.Type   &   Ref.   &   GLON.   &   GLAT.    \\
\hline
 01005+7910   & 0.2  & 0.1   & (25)      &  C   &   (25)   &   B2       &   (25)   &   123.57  &   +16.59   \\
 02229+6208   & 1.67 & 0.09  & (17)      &  C   &   (37)   &   G9Ia     &   (37)   &   133.73  &   +1.50    \\
 04296+3429   & 1.3  & 0.1   & (7, 15, 36)  &  C   &    (27)   &   G0Ia     &   (3)    &   166.24  &   $-$9.05  \\
 05113+1347   & 1.1  & 0.2   & (26, 36, 40) &  C   &    (26)   &   G5I      &   (40)   &   188.86  &   $-$14.29 \\
 05251-1244   & 0.23 & 0.09  & (35)      &  C   &   (42)   &   PN/07f   &   (14)   &   215.21  &   $-$24.28 \\
 05341+0852   & 1.65 & 0.09  & (7, 17, 40)  &  C   &    (26)   &   F5I      &   (15)   &   196.19  &   $-$12.14 \\
 06530-0213   & 1.7  & 0.2   & (36, 40)     &  C   &    (18)   &   F5I      &   (18)   &   215.44  &   $-$0.13  \\
 07134+1005   & 0.4  & 0.1   & (15, 40)     &  C   &    (26)   &   F7 Ie    &   (40)   &   206.75  &   +9.99    \\
 08005-2356   & 0.7  & 0.3   &  (15, 40)     &  O   &   (5)    &   F5I      &   (3)    &   242.36  &   +3.58    \\
 08143-4406   & 0.8  & 0.1   &  (36, 40)     &  C   &   (39)   &   F8I      &   (36)   &   260.83  &   $-$5.07  \\
 08544-4431   & 1.45 & 0.09  &  (29)         &  O   &   (29)   &   F3       &   (29)   &   265.50  &   +0.39    \\
 12175-5338   & 0.25 & 0.09  &  (19, 38)     &  C   &   (46)   &   A9Iab    &   (32)   &   298.30  &   +8.67    \\
 16594-4656   & 2.2  & 0.3   &  (20, 40, 44) &  C   &   (9)    &   B7       &   (44)   &   340.39  &   $-$3.29  \\
 17086-2403   & 0.86 & 0.09  &  (40)         &  C?  &   (10)   &   PN+G5?   &   (34)   &   359.84  &   +8.99    \\
 17097-3210   & 0.06 & 0.05  &  (30)         &  C   &   (45)   &   B9       &   (30)   &   353.36  &   +4.03    \\
 17150-3224   & 0.68 & 0.09  &  (23)         &  O   &   (23)   &   G2I      &   (23)   &   353.84  &   +2.98    \\
 17245-3951   & 1.0  & 0.1   &  (40)         &  O   &   (20)   &   F6I      &   (40)   &   348.81  &   $-$2.84  \\
 17395-0841   & 1.1  & 0.2   &  (11, 40)     &  O?  &   (10)   &   PN+G     &   (11)   &   17.02   &   +11.10   \\
 17423-1755   & 1.13 & 0.09  &  (11, 40)     &  O   &   (41)   &   Be       &   (11)   &   9.36    &   +5.78    \\
 17436+5003   & 0.24 & 0.09  & (19)      &  O   &   (4)    &   F3 Ib    &   (48)   &   77.13   &   +30.87   \\
 18025-3906   & 1.15 & 0.09  & (22, 40)     &   O   &   (24)   &   G1I      &   (40)   &   353.27  &   $-$8.72  \\
 18062+2410   & 0.6  & 0.2   & (33, 40)     &   O   &   (33)   &   B1 I     &   (33)   &   50.67   &   +19.79   \\
 HD 172324    & 0.03 & 0.02  & (6)       &  O   &   (1)    &   B9Ib     &   (1)    &   66.18   &   +18.58   \\
 19114+0002   & 0.60 & 0.05  & (19, 40)     &   O   &   (4)    &   G5 Ia    &   (3)    &   35.62   &   $-$4.96  \\
 19200+3457   & 0.3  & 0.1   & (11)      &  O?  &   (11)   &   B        &   (11)   &   67.57   &   +9.51    \\
 19386+0155   & 1.05 & 0.09  & (2)       &  O   &   (28)   &   F5I      &   (31)   &   40.51   &   $-$10.09 \\
 19500-1709   & 0.37 & 0.09  & (19)      &  C   &   (47)   &   F2-6 Ia  &   (3)    &   23.98   &   $-$21.04 \\
 20000+3239   & 1.6  & 0.1   & (15, 26)     &   C   &   (21)   &   G8Ia     &   (21)   &   69.68   &   +1.16    \\
 20462+3416   & 0.38 & 0.09  & (8,  43)     &   O   &   (8)    &   B1.5     &   (8)    &   76.60   &   $-$5.75  \\
 22023+5249   & 0.52 & 0.09  & (12)      &  O   &   (13)   &   B        &   (12)   &   99.30   &   $-$1.96  \\
 22223+4327   & 0.2  & 0.1   & (15, 26)     &   C   &   (26)   &   G0 Ia    &   (15)   &   96.75   &   $-$11.56 \\
 22272+5435   & 0.9  & 0.2   & (15)      &  C   &   (16)   &   G5 Ia    &   (15)   &   103.35  &   $-$2.52  \\
 23304+6147   & 1.4  & 0.2   & (15, 36)     &   C   &   (15)   &   G2 Ia    &   (15)   &   113.86  &   +0.59    \\
\hline
\end{tabular}
\resizebox{1\textwidth}{!}{
\begin{tabular}{l l l l}\\
(1) \citet{Arellano01}      &   (13) \citet{Gauba04}    & (25) \citet{Klochkova02}&     (37) \citet{Reddy99}  \\
(2) \citet{Arkhipova00}     &   (14) \citet{Heap87}     & (26) \citet{Kwok95}     &     (38) \citet{Reed96}  \\
(3) \citet{Bakker97}        &   (15) \citet{Hrivnak95}  & (27) \citet{Kwok99}     &     (39) \citet{Reyniers04} \\
(4) \citet{Bujarrabal92}    &   (16) \citet{Hrivnak91}  & (28) \citet{Lewis00}    &     (40) \citet{Suarez06} \\
(5) \citet{Desmurs02}       &  (17) \citet{Hrivnak99a}      & (29) \citet{Maast03}    &     (41) \citet{Te_Lintel_Hekkert91} \\
(6) \citet{Fernie83}        &  (18) \citet{Hrivnak03}   & (30) \citet{Malfait98}  &         (42) \citet{Torres-Peimbert80} \\
(7) \citet{Fuji02}      &  (19) \citet{Hrivnak89}   & (31) \citet{Meixner99}  &     (43) \citet{Turner84} \\
(8) \citet{Garcia-Lario97}  &  (20) \citet{Hrivnak99b}  & (32) \citet{Oudmaijer92}&     (44) \citet{Steene03} \\
(9) \citet{Garcia-Lario99}  &  (21) \citet{Hrivnak00}   & (33) \citet{Parthasarathy00a} &   (45) \citet{Van_der_Veen89} \\
(10) Garc\'\i a-Lario (priv.comm.)& (22) \citet{Hu93a}      & (34) \citet{Parthasarathy00b} &   (46) \citet{Van_Winckel97} \\
(11) \citet{Gauba03a}       &   (23) \citet{Hu93b}      & (35) \citet{Pottasch04} &     (47) \citet{Van_Winckel00} \\
(12) \citet{Gauba03b}       &    (24) \citet{Hu94}      & (36) \citet{Reddy96} &        (48) \citet{Volk89} \\
\end{tabular}
}
\end{table*}


In this work we present the first systematic survey to detect the
presence of circumstellar DBs (DCBs) in a carefully selected sample of galactic
post-AGB stars\footnote{The sample also includes three very young planetary
nebulae, which are here also considered post-AGB stars in a broad sense.}. The
goal is to perform a detailed analysis of the differential properties observed
in the DBs associated to post-AGB stars in comparison with the standard DBs
observed towards reddened, early-type field stars (where these bands
are expected to be essentially of interstellar origin).

To perform this task we have studied the intensity of 9 of the strongest 
absorption features identified as DBs in the spectral range 4000--10000~\AA\ 
using high resolution optical 
spectroscopy. The comparison of the properties observed in carbon-rich and
oxygen-rich post-AGB stars is used to test the carbon-rich nature of the DB
carrier(s) and determine which of the DBs detected are most probably of
circumstellar origin (if any).

In Section~\ref{section:observation} we describe the observations made and the
data reduction process. The strategy followed in our analysis is presented in
Section~\ref{section:relation}. The main results are discussed in Section 4 as
a function of various observational parameters. Finally, the conclusions are
presented in Section~\ref{section:conclusions}.

\begin{figure}[t!]
\begin{center}
\includegraphics[width=0.9\columnwidth]{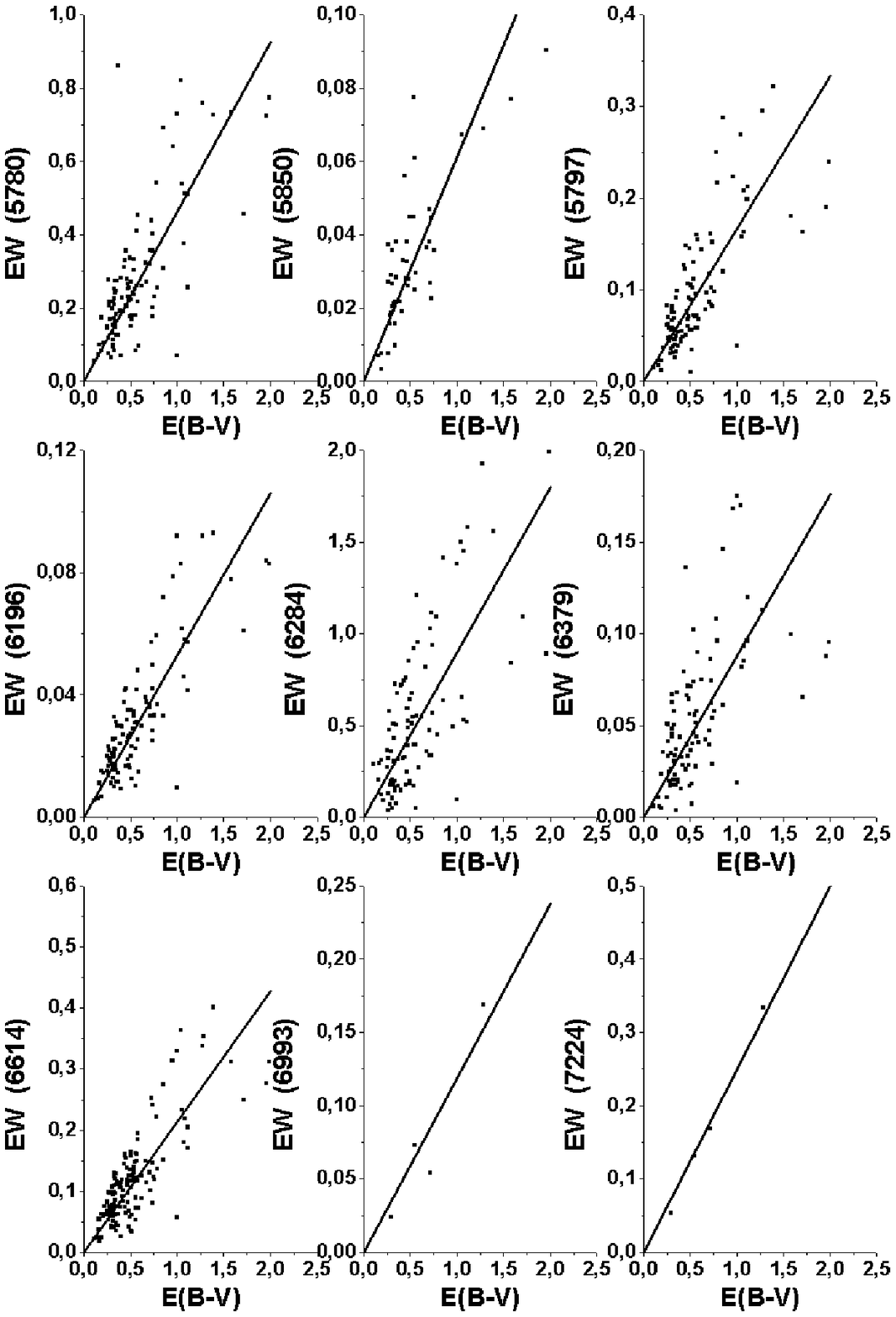}
\caption{
Equivalent width measurements taken from the literature for field stars,
plotted as a function of $E(B-V)$. Table~\ref{tb:caracdbs} gives slopes
$EW$/$E(B-V)$ and correlations \emph{r} for the linear fits.}
\label{fig:thor_02}
\end{center}
\end{figure}

\section{Observations and Data Reduction}
\label{section:observation}

The high-resolution Echelle spectroscopic data analysed in this paper were
taken using a wide variety of instruments and telescopes over the period
1993-2003. Originally, these observations were carried out for chemical
abundance analysis purposes and they correspond to observations performed using
the Utrecht Echelle Spectrograph (UES) at the William Herschel Telescope (WHT
4.2m) and the High Resolution Spectrograph (SARG) at the Telescopio Nazionale
Galileo (TNG 3.58m), both in the Spanish Observatorio del Roque de los
Muchachos (La Palma, Spain); the UV-Visual Echelle Spectrograph (UVES)
installed at the Very Large Telescope-U2 (VLT 8m) in Paranal Observatory
(Chile); and the ESO Multi-Mode Instrument (EMMI) at the New Technology
Telescope (NTT 3.5m) and the Fiber-fed Extended Range Optical Spectrograph
(FEROS) at the ESO 1.52m telescope in La Silla Observatory (Chile).

The spectra obtained cover a wide wavelength range (usually from 4000 to
10000~\AA) with a resolving power in the range 50000$-$100000 at 5500~\AA,
depending on the instrument set-up. The exposure times are variable, depending
on the brightness of the source, but typically of $\sim$~30 min, leading to a
signal-to-noise of 20$-$200 over the spectral range considered.

The two-dimensional spectra were reduced following the standard procedure for
echelle spectroscopy using IRAF astronomical routines. The process includes:
identification of bad pixels, bias determination and scattered light
subtraction, flat-field correction, order extraction and wavelength
calibration. For the DBs at 6284, 6993 and 7224 $\AA$, strongly affected by
terrestrial features, the spectrum of \object{HD 172324} (B9Ib), one of the
sample stars very little affected by extinction, was used as divisor to remove
the telluric absorption lines\footnote{Note that the initial strategy was to
use the spectrum of a hot, rapidly rotating star observed on the same night for
this purpose, but it was found that the latter showed faint but detectable DBs
in its spectrum which did not allow us to perform this correction properly. In
contrast, the 6284, 6993 and 7224~\AA\ DBs were found to be totally absent in
\mbox{\object{HD 172324}}.} (see Figure~\ref{fig:mezcla}).

\begin{figure}[ht!]
\begin{center}
\includegraphics[width=.9\columnwidth]{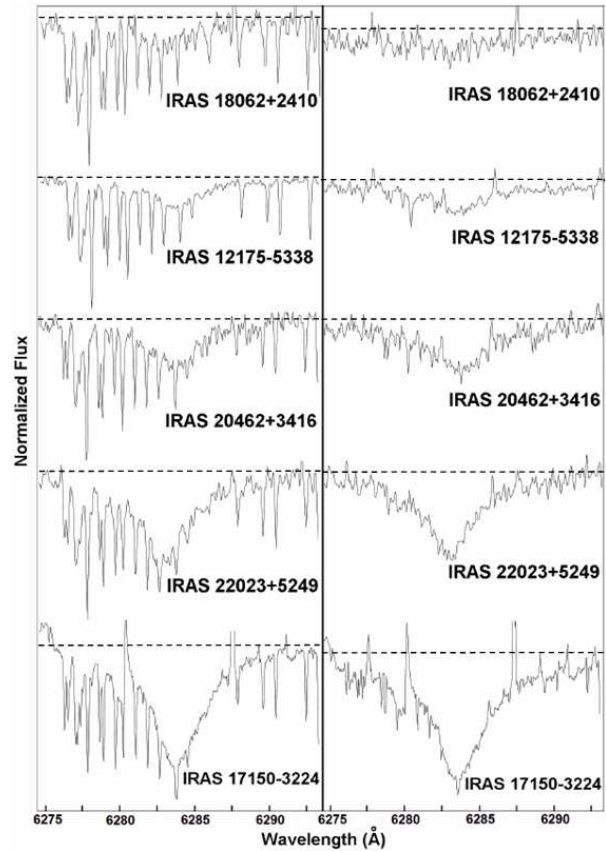}
\caption{Sample spectra showing the region around the DB at 6284~\AA,
before (left panel) and after (right panel) removal of the
telluric lines using as the telluric divisor \object{HD 172324}.
Dotted lines indicate the continuum level adopted
in each case.} \label{fig:telluric}
\end{center}
\end{figure}

\section{Selection of the sample}\label{section:relation}

In order to make a systematic study of the presence of DCBs in post-AGB stars,
we carefully selected a sample of 33 sources from the literature trying to
cover as much as possible a wide variety of observational properties such as
the chemistry of the circumstellar envelope (carbon-rich and oxygen-rich) or
the spectral type of the central stars\footnote{Stars with spectral types later
than G- were discarded for the analysis, as their continuum is dominated by the
presence of molecular bands, which makes the identification of DBs a very
difficult task.}. Priority was given to sources located at high galactic
latitudes showing a strong colour excess $E(B-V)$ (as these sources are
most likely dominated by circumstellar extinction) and to those for which a
high radial velocity has been reported in the literature (since this may later
facilitate the identification of spectral features of circumstellar origin).

The main objective is to understand whether systematic differences are detected
which depend on one (or more) of the above observational parameters.

The list of stars selected for analysis, most of them IRAS sources belonging
to the GLMP catalogue of post-AGB stars \citep{Garcia-Lario97, Suarez04},
is displayed in Table~\ref{tb:observations}, where a summary description of
the observations made is presented. In Table~\ref{tb:referencias}, additional
information is given on the sources included in our observing programme. This
includes the colour excess $E(B-V)$, dominant chemistry (carbon-rich or
oxygen-rich), spectral type and galactic coordinates (GLON, GLAT), as well as
the bibliographic references from where this information was extracted.

The spectral regions corresponding to 9 different DBs which are among the
strongest ones reported in the literature have been investigated in detail for
each of the sources included in our sample. Table~\ref{tb:caracdbs} lists the
accurate wavelengths ($\lambda_0$) corresponding to each of these features,
taken from \citet{Galazutdinov00}, as well as their central depth $A_C$ and
sensitivity to the extinction, measured as $EW/E(B-V)$, observed toward the
star \object{HD 183143} (B7\,I; \mbox{$E(B-V)$ = 1.28 mag)}, which is usually
taken as the prototype star in the analysis of DIBs \citep{Herbig95}. All the
DBs selected for analysis are within the optical domain, and they are referred
to in the literature as the {\it 5780, 5797, 5850, 6196, 6284, 6379, 6614,
6993} and {\it 7224~\AA\ features}. Other well known DBs at 4430 and 6177~\AA\
are even stronger than the selected ones, but they have been discarded for
study because of the difficulty to detect their extremely broad (and relatively
shallow) profiles (FWHM $>$ 17~\AA; $A_c < 0.1$) in our high resolution spectra.


\begin{table}[t!]
\caption{
Main characteristics of the selected DBs (central depth $A_C$
and normalised equivalent width $EW$/$E(B-V)$), as measured
towards the prototype star HD 183143 \citep{Herbig95} (cols.~3 and 4).
The reference wavelengths are taken from \citet{Galazutdinov00} (col.~2).
The equivalent width per extinction unit (this work) derived from
published data from \citet{Jenniskens94b}, \citet{Weselak01},
\citet{Thorburn03} and \citet{Megier05} is given for each DIB
in col.~5, with corresponding correlation coefficients in col.~6.
}
\label{tb:caracdbs} \centering
\begin{tabular}{c c c c c l} \hline \hline
    &       & \multicolumn{2}{c}{HD 183143}         &\multicolumn{2}{c}{this work}\\
\cline{3-4}\cline{5-6}
DB      & $\lambda_{0}$ &     $A_c$    &   $EW$/$E(B-V)$  &   $EW$/$E(B-V)$ &   r         \\
(\AA)   &    (\AA)      &          &  (\AA/mag) &   (\AA/mag)   &         \\
\hline
5780    &   5780.37     &      0.32    &   0.63     &   0.46    &   0.74      \\
5797    &   5796.96     &      0.20    &   0.19     &   0.17    &   0.73      \\
5850    &   5849.80     &      0.069   &   0.06     &   0.061   &   0.75      \\
6196    &   6195.96     &      0.084   &   0.06     &   0.053   &   0.82      \\
6284    &   6283.85     &      0.32    &   1.5      &   0.90    &   0.69      \\
6379    &   6379.29     &      0.10    &   0.096    &   0.088   &   0.62      \\
6614    &   6613.56     &      0.24    &   0.29     &   0.21    &   0.80      \\
6993    &   6993.18     &      0.14    &   0.14     &   0.12    &   0.95      \\
7224    &   7224.00     &      0.21    &   0.29     &   0.25    &   0.99      \\
\hline
\end{tabular}
\begin{list}{}{}
\item $A_c$ = 1 - F($\lambda_{0}$) / F(continuum)
\end{list}

\end{table}


\section{Discussion}

\subsection{DB strength vs. extinction in field stars}\label{section:fieldstars}

Although it is generally accepted that there is a tight correlation between the
equivalent width of DBs and the value of $E(B-V)$ in field stars dominated by
interstellar reddening, the available results in the literature generally cover
only the stronger DBs ($\lambda\lambda$~4430, 5780, 5797 and 6284). In
addition, they are usually based on old data obtained with poor spectroscopic
resolution (sometimes, even below R~=~1000). Unfortunately, studies covering
other DBs and/or based on high resolution spectroscopy are scarce. Prior to
derive any conclusion on the existence (or not) of a similar correlation
between DB strength and extinction in our sample of post-AGB stars it is, thus,
necessary to establish this dependency for each of the 9 DBs observed
toward field stars.

For this purpose, we have re-derived ourselves these correlation parameters
using a large number of early-type stars taken from
\citet{Weselak01,Thorburn03} and \citet{Megier05}, for which accurate DB
strength measurements are available covering a wide range of extinction values.
Weselak et al.'s sample contains 41 stars observed at R$\sim$64000, for which
equivalent width data are available corresponding to the DB centred at
6614~\AA. Thorburn et al.'s sample comprises 53 stars observed with spectral
resolution R$\sim$38000, that were originally used to study the DBs centred at
5780, 5797, 6196, 6284, 6379 and 6614~\AA. Finally, Megier et al.'s sample
comprises 49 stars observed with spectral resolution R$\sim$64000, that were
originally used to study the DBs centred at 5780, 5797, 5850, 6196, 6284, 6379
and 6614~\AA. Four additional stars taken from \citet{Jenniskens94b}, observed
at R$\sim$20000, were also used to derive the correlation parameters associated
to the DBs centred at 6993 and 7224~\AA, for which measurements are much more
scarce in the literature. The sources included in these four samples are
expected to follow a behaviour representative of field stars affected only by
interstellar extinction. Figure \ref{fig:thor_02} shows all the equivalent
width measurements used in our analysis, plotted as a function of the
interstellar extinction, measured as $E(B-V)$.

For each DB under analysis we have applied a linear fit to the data available.
We have also imposed the condition $EW$ = 0 for $E(B-V)$ = 0, i.e.:
$EW$ = $a \cdot$ $E(B-V)$, where \emph{a} is a constant that represents the
equivalent width per extinction unity. In practice this is equivalent to assume
that there is a direct link between the DB carrier(s) and the material which is
responsible for the extinction observed in the ISM. The fits obtained represent
the DB strength expected as a function of the colour excess for any given
source in which interstellar reddening is the dominant contributor to the
overall extinction. These are represented by solid lines in Figure
\ref{fig:thor_02}. The slopes ($EW$/$E(B-V)$) and correlation coefficients
\emph{r} of the linear fits are given in Table~\ref{tb:caracdbs}.

As we can see, a reasonable correlation between equivalent width and $E(B-V)$
is always found, although the dispersion is in some cases considerable.
The new results obtained are in agreement as well with those derived
for the prototype star
\mbox{\object{HD 183143}} by \citep{Herbig95} although the 5780 and 6284 DIBs
are significantly stronger toward the latter, probably due to local environmental
conditions.
These results provide confidence to proceed with the study
of the post-AGB stars in our sample, based on the
assumption that the above values can be taken as references for the
subsequent analysis.

\begin{figure*}[ht!]
\begin{center}
\includegraphics[width=12cm]{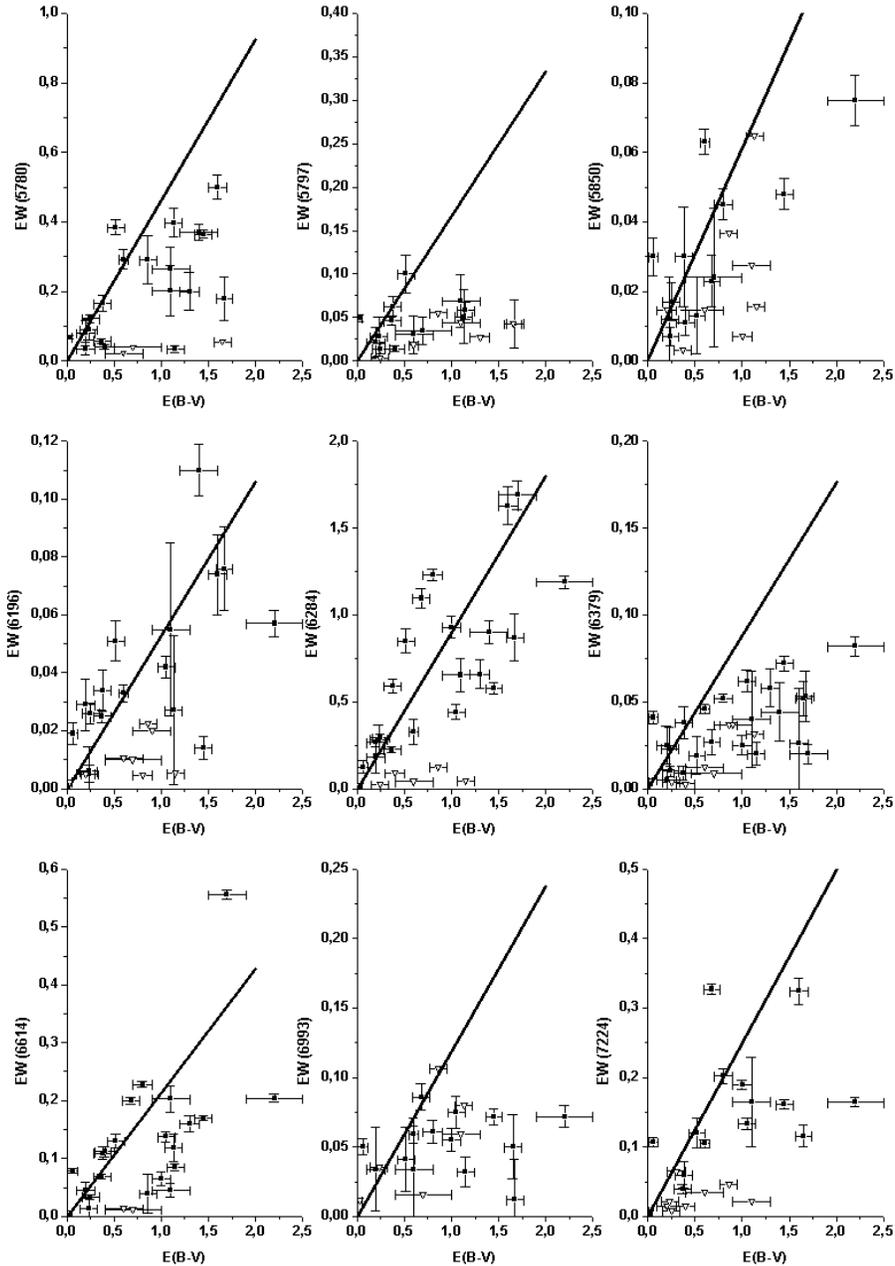}
\caption{Equivalent width (in \AA) of the 9 DBs selected for analysis as a function of
$E(B-V)$ for the post-AGB stars in the sample. Solid lines correspond to the fits
derived in this paper for field stars dominated by interstellar extinction
(see also Sect.~\ref{section:fieldstars}).
The inverted triangles represent upper limits.}
 \label{fig:intensidad1}
\end{center}
\end{figure*}

\subsection{DB strength vs. extinction in post-AGB stars}

In Table~\ref{tb:intensidades} (Online only) we show the equivalent width of each of the 9
DBs considered in our analysis for every post-AGB star in the sample, as
determined from the available high resolution spectra. Note that three of these
bands are strongly affected by telluric contamination, namely those centred at
6284, 6993 and 7224~\AA. For these features, a careful removal of the telluric
component was performed prior to the determination of their $EW$ (see Figures
\ref{fig:mezcla} and \ref{fig:telluric}; Sect.~\ref{section:observation}).
For the other DBs, the measurements were performed directly on the normalised spectra.
We note here that in none of the spectra we found evidence for DB features
in emission.

The resulting values are plotted in Figure~\ref{fig:intensidad1} as a function
of the colour excess \mbox{$E(B-V)$}. The values of $E(B-V)$ which were used to
produce this figure are directly taken from the literature (see
Table~\ref{tb:referencias}) or estimated from the available information on
spectral type and photometry by comparing the observed B and V magnitudes with
the intrinsic $B-V$ colours expected for stars of the same spectral type and
luminosity class~I\footnote{A luminosity class I is adopted because this is the
class corresponding to low-gravity stars, but note that post-AGB stars only
look like super-giants, but they are not genuine population I massive super-giant
stars \citep{Fitzgerald70}. Values quoted by different authors are generally
in good agreement, and the discrepancies found, when relevant, are reflected in
the associated errors provided in Table \ref{tb:referencias}.}.

Figure~\ref{fig:intensidad1} shows the overall results obtained for the 9 DBs
studied. In general we find that the equivalent width of the observed features
seems to be still correlated with the value of $E(B-V)$. However, in contrast
to the results obtained for the field stars (shown in
Figure~\ref{fig:thor_02}), this correlation is now very weak in some cases and
we identify a much larger number of outliers.

Usually, for a given extinction $E(B-V)$, the measured equivalent widths in
post-AGB stars are well below the expected values. Only a subset of sources
follow exactly the same behaviour observed in field stars. We interpret this
result as the consequence of the absence (or at least the under-abundance) of
the DB carriers in the circumstellar envelopes of most of these post-AGB
stars, but further analysis is needed to confirm that there is no other
alternative explanation.

\begin{table*}[ht!]
\caption{
Line-of-sight properties for the observed post-AGB stars.
DCS = dominated by circumstellar reddening; HV = high radial velocity;
CS1 designated DCS based on lat/long vs reddening, excluding those
that are discarded based on interstellar (IS) reddening estimate.
CS2 indicates DCS found by estimating the upper limit to the IS reddening.
Principal extinction estimates are given with rescaling of disk / spiral
component (col. 7).
No-rescaling estimates are given (col. 8) in those cases for which the extinction
estimate is significantly higher than with use of rescaling.
The angular scale of COBE data is 0.35$\degr$ $\times$ 0.35$\degr$ and this
dust extinction model therefore only gives mean extinction estimates.
Erroneous rescaling factors can arise for directions toward strong extra-galactic sources
such as M31, M33, SMC and LMC as well as toward peculiar galactic regions such as Orion
and the Rho Ophiuchus complex.
Also, lines of sight corresponding to arm tangents may have large systematic errors.
Distance estimates (references in col. 4) and corresponding model extinctions (converted
to $E(B-V)$) are given when available in Cols.~3 \& 5).
The maximum model reddening (with and without re-scaling) and the corresponding distance
in the target direction are given in cols. 7, 8 and 6, respectively.
Col. 9 gives the resulting lower limit for the circumstellar reddening.
The final column (10) indicates when the target is dominated by circumstellar reddening (CS1 or CS2)
and/or is a high velocity target (HV).
}
\centering
\label{tb:extinction}
\begin{tabular}{llllllllll}\hline\hline
    NAME      & $E(B-V)$     & d       & Ref.     &$E(B-V)$IS& d$_{\rm max}$&\multicolumn{2}{c}{$E(B-V)$IS} & $E(B-V)$-CS     & DCS / HV  \\
          & observed     & kpc     &          &          & kpc      &scaling   & no scaling& min              &     \\\hline
 01005+7910   & 0.2  $\pm$ 0.1   & $~$ 3   &          &  0.11    &  1.7     & 0.11     &       &              &      \\
 02229+6208   & 1.67 $\pm$ 0.09  & $>$ 2.2 &          &  0.60    &  5       & 0.99     &       & 0.68         & CS2  \\
 04296+3429   & 1.3  $\pm$ 0.1   & 3.5     &          &  0.23    &  3.3     & 0.23     &       & 1.07         & CS1  \\
              &                  & 5.4, 5  & 2, 9     &  0.23    &      &          &       &              &      \\
 05113+1347   & 1.1  $\pm$ 0.2   & 5       & 9        &  0.14    &  2.1     & 0.14     &       & 0.96         & CS1  \\
 05251-1244   & 0.23 $\pm$ 0.09  &     &          &      &  1       & 0.18     &       &              &      \\
 05341+0852   & 1.65 $\pm$ 0.09  & 10      & 9        &  0.16    &  2.4     & 0.16     &       & 1.49         & CS1  \\
 06530-0213   & 1.7  $\pm$ 0.2   & 3       & 9        &  0.36    &  5       & 0.41     &       & 1.29         & CS2  \\
 07134+1005   & 0.4  $\pm$ 0.1   &     &          &          &  2.8     & 0.024    & 0.14      & 0.38 / 0.26      & HV / CS2 \\
 08005-2356   & 0.7  $\pm$ 0.3   &     &          &      &  4       & 0.18     & 0.27      & 0.52 / 0.43      & CS2  \\
 08143-4406   & 0.8  $\pm$ 0.1   & 4       & 9        &  0.55    &  5.8     & 0.66     &       &              &      \\
 08544-4431   & 1.45 $\pm$ 0.09  &     &          &      &  6       & 0.99     &       & 0.46         & CS2?     \\
 12175-5338   & 0.25 $\pm$ 0.09  &     &          &      &  3       & 0.24     &       &              &      \\
 16594-4656   & 2.2  $\pm$ 0.3   & 2.6     & 10,11    &      &  7       & 2.1      &       &              & CS (*)   \\
 17086-2403   & 0.86 $\pm$ 0.09  & 6       & 9        & 0.56     &  2       & 0.56     &       & 0.3          & HV (CS2?)\\
 17097-3210   & 0.06 $\pm$ 0.05  & (0.2)   &          & (0.08)   &  5       & 0.95     &       &              &      \\
 17150-3224   & 0.68 $\pm$ 0.09  & $<$ 18 (2) & 9     & (0.9)    &  5       & 1.8      &       &              &      \\
 17245-3951   & 1.0  $\pm$ 0.1   &     &          &      &  5       & 1.0      & 1.5       &              & HV   \\
 17395-0841   & 1.1  $\pm$ 0.2   &     &          &      &  2       & 0.87     &       & 0.23         & HV \\
 17423-1755   & 1.13 $\pm$ 0.09  & 3.2 -- 3.7 &       &   0.6    &  4       & 0.60     &       & 0.53         & CS1 / HV \\
 17436+5003   & 0.24 $\pm$ 0.09  & $~$ 1.2    &       & 0.03     &  0.9     & 0.026    & 0.064     & 0.18         & CS1  \\
 18025-3906   & 1.15 $\pm$ 0.09  &     &          &      &  2       & 0.23     & 0.34      & 0.92 / 0.81      & CS1 / HV \\
 18062+2410   & 0.6  $\pm$ 0.2   & 4.5 -- 5.3 &       & 0.14     &  1.4     & 0.14     &       & 0.46         & CS1 / HV \\
 HD 172324    & 0.03 $\pm$ 0.02  &     &          &      &  1.5     & 0.056    & 0.11      &              &      \\
 19114+0002   & 0.60 $\pm$ 0.05  &     1.5    & 4     &   0.32   &  5       & 0.54     & 0.62      &              & HV   \\
 19200+3457   & 0.3  $\pm$ 0.1   &     &          &      &  2.9     & 0.15     & 0.22      &              &      \\
 19386+0155   & 1.05 $\pm$ 0.09  &     &          &      &  2.9     & 0.35     &       & 0.80         & CS1  \\
 19500-1709   & 0.37 $\pm$ 0.09  & $>$ 4   & 12       &      &  1.0     & 0.20     &       &              &      \\
 20000+3239   & 1.6  $\pm$ 0.1   & (5,8)   &          &(1.0,1.7) & $>$10    &$>$1.7    &       & ?            &          \\
 20462+3416   & 0.38 $\pm$ 0.09  &     &          &      &  4       & 0.25     & 0.38      &              & HV   \\
 22023+5249   & 0.52 $\pm$ 0.09  & 3.3 -- 3.9 &       & 0.47,0.50&  5       & 0.52     & 0.55      &              & HV   \\
 22223+4327   & 0.2  $\pm$ 0.1   &     &          &      &  2.6     & 0.21     &       &              &      \\
 22272+5435   & 0.9  $\pm$ 0.2   &  1.6,2.7   & 8     & 0.26,0.39&  4.5     & 0.47     &       & 0.43         & CS2  \\
 23304+6147   & 1.4  $\pm$ 0.2   &  4.7, 5    & 5, 9  &  0.64    &  4.5     & 0.64     &       & 0.76         & CS2  \\
\hline
\end{tabular}
\begin{tabular}{p{\textwidth}}
\ [1] \citet{Reddy99}; [2] \citet{Klochkova99}; [3] \citet{Garcia-Lario99}; [4] \citet{Zacs99,Zacs99b};
\ [5] \citet{Klochkova00}; [6] \citet{Bakker97}; [7] Hrivnak \& Kwok 1991; [8] \citet{Woodsworth90}; [9] \citet{Reddy96};
\ [10] \citet{Hrivnak00}; [11] \citet{Steene03}; [12] \citet{Clube04}, A(V)=1.2.
(*) \object{IRAS 16594-4656} is a well known bi-polar PPN which is thought to have
a high internal extinction (see also Sect.~\ref{sec:extinction}).
\end{tabular}
\end{table*}

\subsection{Interstellar vs. circumstellar extinction}\label{sec:extinction}

In order to determine whether our preliminary hypothesis is consistent with the
measurements here presented, it is necessary to take into account that in
general, the overall extinction observed towards a given source in the sky is
the result of the combined effect of the contribution coming from the ISM and
of the internal extinction produced in the circumstellar shell. Making this
distinction is generally not important in field stars, since for them the
latter contribution is negligible. However, for the evolved stars in our sample
the situation is completely different as, in many cases, the observed reddening
is almost exclusively of circumstellar origin.

Disentangling interstellar versus circumstellar extinction for a given source
is a very difficult task, if we need to rely only on the available
observations. The only option we have is to use a statistical approach to 
estimate whether the observed extinction corresponds preferentially to one or
another component.

\begin{figure}[t!]
\begin{center}
\includegraphics[width=\columnwidth]{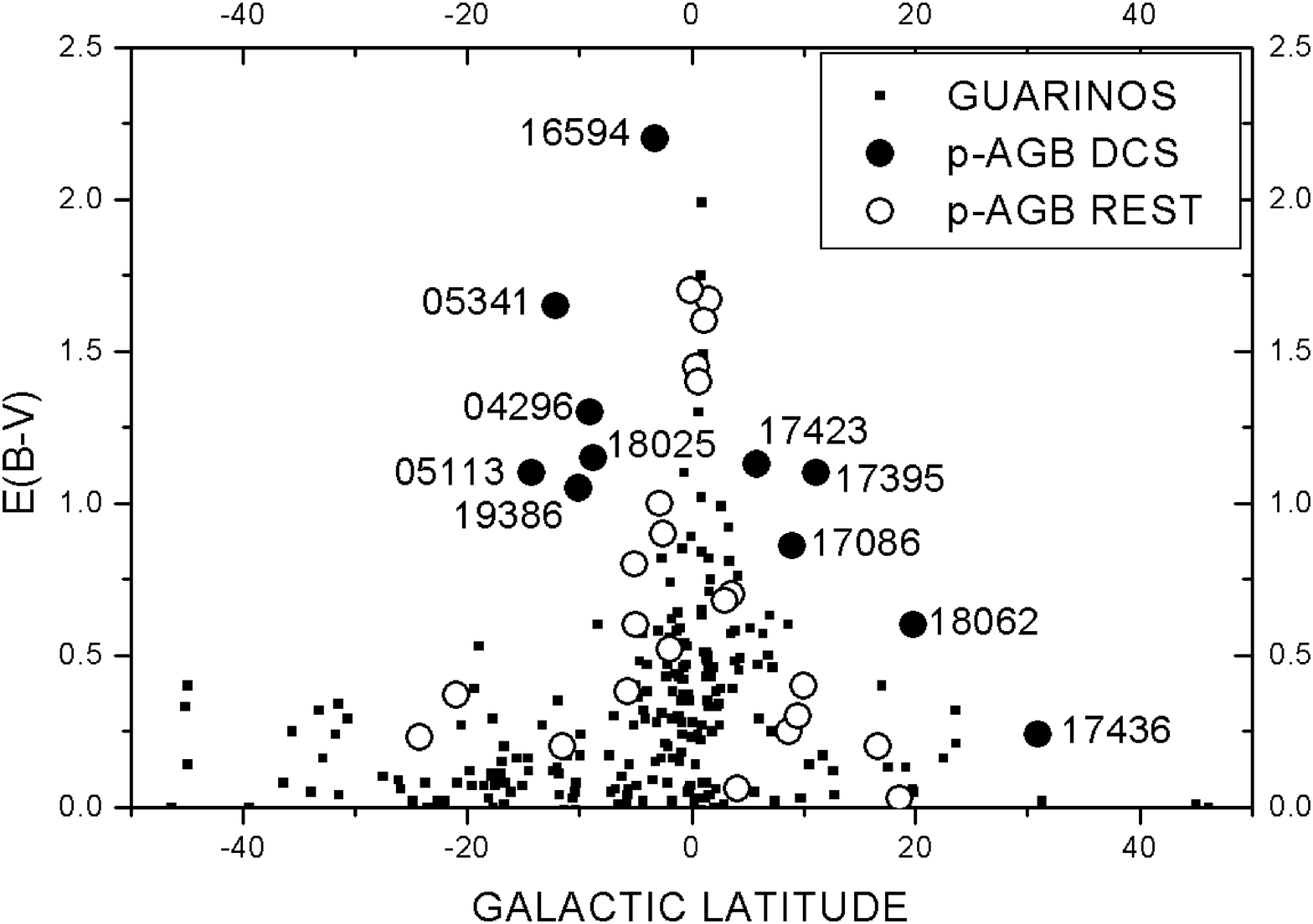}
\caption{$E(B-V)$ versus Galactic latitude distribution of the
post-AGB stars in the sample (circles) and for reference stars
taken from \citet{Guarinos88a, Guarinos88b, Guarinos97} (small squares).
Post-AGB stars dominated by circumstellar extinction (DCS-type) are
indicated by filled circles and they are labeled with their IRAS
name.}
\label{fig:seleccion}
\end{center}
\end{figure}

For this we have represented in Figure~\ref{fig:seleccion} the $E(B-V)$ versus
galactic latitude distribution of the post-AGB stars in our sample and compared
this distribution with that shown by field stars taken from the catalogue of
Guarinos \citep{Guarinos88a,Guarinos88b, Guarinos97}. This catalogue, also used
for the study of DB strengths, contains observations of 270 early-type field
stars homogeneously distributed along the Galactic Plane (but excluding the
Galactic Bulge), located at a variety of galactic latitudes and for which the
value of $E(B-V)$ has previously been determined.

Figure \ref{fig:seleccion} shows clearly that a subsample of post-AGB stars are
clear outliers in this plot. This indicates that the reddening excess in these
stars must be circumstellar in origin. Other sources, however, show a
relatively small reddening fully compatible with the values observed in field
stars located at the same galactic latitude.


Based on this analysis, we have divided our sample of post-AGB stars in two
groups according to whether the overall extinction observed is more likely to
be dominated by the circumstellar contribution (DCS-type stars; filled circles
in Figure \ref{fig:seleccion}; CS1 in Table~\ref{tb:extinction}) or just
consistent with the interstellar extinction expected according to its Galactic
Latitude (rest of stars; indicated by open circles in Figure
\ref{fig:seleccion}).

Note that the above classification is very rough and that it just considers a
star as belonging to the DCS group if it shows a relative large reddening
excess with respect to the nominal value expected from its galactic location.
Stars in which there is only a moderate (although possibly significant)
contribution from the circumstellar shell to the observed extinction may have
escaped detection. This means that the group of stars not classified in the DCS
group may still contain sources in which the circumstellar contribution to the
observed reddening is not negligible, and vice versa, those classified
DCS may still contain a significant interstellar dust contribution.

In order to estimate an upper limit to the contribution of the
interstellar reddening to the total reddening we use the Galactic 3D-extinction
model map by \citet{Drimmel03} that gives the \emph{mean} visual extinction as a
function of sky (galactic) coordinates and distance. The extinction has a
projected resolution of 0.35$\degr$ $\times$ 0.35$\degr$ (this is set by the
COBE map which is used to re-scale the extinction in order for the model to
reproduce correct far-infrared flux). It is evident that any small scale
structure (including the circumstellar contribution of the target star) is
washed out in these estimates, which nevertheless give us information on the
\emph{global} spatial distribution of dust in specific directions.  We take
distance estimates from literature where possible (column~3 in
Table~\ref{tb:extinction}) and/or extract the maximum extinction (column~7) and
the corresponding distance (column~6) for a particular line-of-sight. For high
latitudes the extinction versus distance curve flattens rapidly, within a few
kpc. These (upper limit) estimates for the interstellar visual extinction, converted
to reddening by dividing by the canonical value for $R_V$~=~3.1, can then be
subtracted from the total observed reddening and thus yield (lower limit)
estimates for the circumstellar reddening (column~9). Caution is needed when
applying these  results to derive extinction values for individual sightlines.
Notwithstanding, we can now review and improve our (statistical)
classification  based on reddening versus latitude.

We consider all targets that are estimated to have a (lower limit)
circumstellar reddening contribution higher than 50\% of the total observed
reddening to be of DCS-type (CS2 in Table~\ref{tb:extinction}).
From the 11 DCS-type targets previously selected,
on basis of their latitude reddening excess, we find that 9 satisfy this
criterion. For the other 2 targets we can attribute a high fraction of the
total observed reddening to interstellar dust.

Among the 22 IS (interstellar) reddening dominated targets selected above
via their latitude we find 7 targets with a non-negligible
circumstellar reddening $E(B-V)$$_{\rm CS} \geq 0.4$~mag
(Table~\ref{tb:extinction}; indicated CS2).

\mbox{\object{IRAS 02229+6208}} and \mbox{\object{IRAS 08544-4431}}
are special cases because these lines-of-sight show both a strong interstellar
and circumstellar dust contribution.

\subsection{DB strength in DCS-type post-AGB stars}

Following the above criteria, we find that for 17 (out of 33) stars in the sample
a significant fraction of the observed reddening is due to the presence of
circumstellar dust (i.e. DCS-type), while for the remaining targets the colour
excesses are expected to be predominantly due to interstellar dust.

\begin{figure}[t]
\begin{center}
\includegraphics[width=\columnwidth]{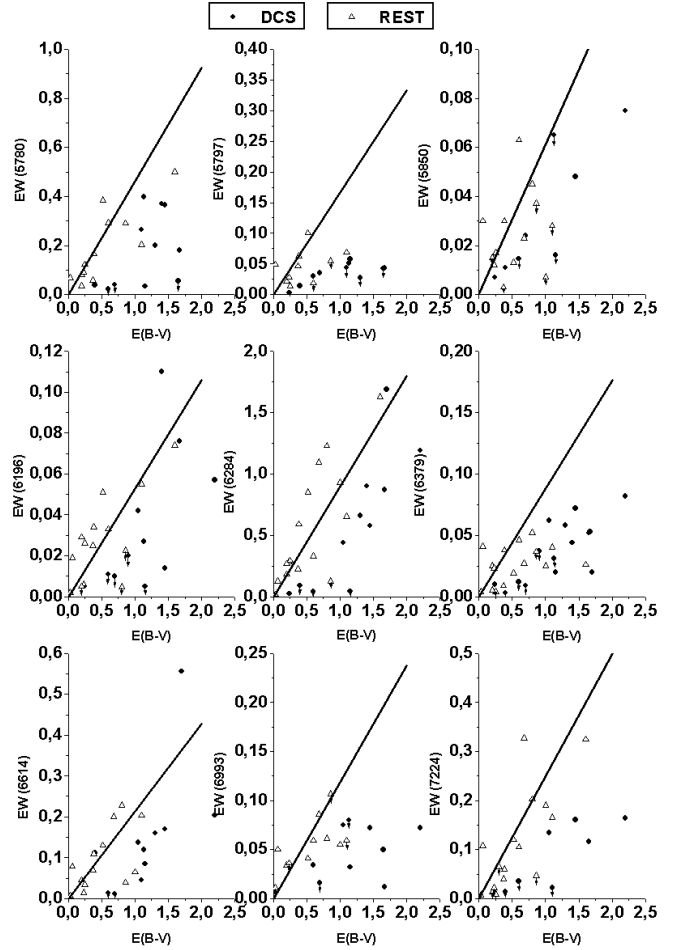}
\caption{Equivalent width of the 9 DBs selected for analysis as a function of
$E(B-V)$ for the post-AGB stars in the sample in which filled circles
represent the subsample of post-AGB stars dominated by circumstellar
extinction (DCS-type) and open triangles the rest of stars in the sample.
 Solid lines again correspond to the fits derived in this paper for field
stars dominated by interstellar extinction. The arrows indicate upper limits.}
 \label{fig:intensidad2}
\end{center}
\end{figure}

A comparison of the DB strengths measured in stars belonging to each of the two
groups considered above with those found in reference field stars is presented
in Figure \ref{fig:intensidad2}, where we can see that there is a general trend
for the stars with a dominant circumstellar extinction contribution (DCS-type)
to show much weaker DB strengths compared to the rest of post-AGB stars in the
sample. In the most extreme cases, there are stars in this group affected by a
large overall extinction in which surprisingly some DBs are completely absent,
and only upper limits to their equivalent width can be reported.
In contrast, we find a significant number of sources among the rest of
post-AGB stars in the sample showing DB strengths fully in agreement with those
observed in field stars.

\begin{figure}
\begin{center}
\includegraphics[width=\columnwidth]{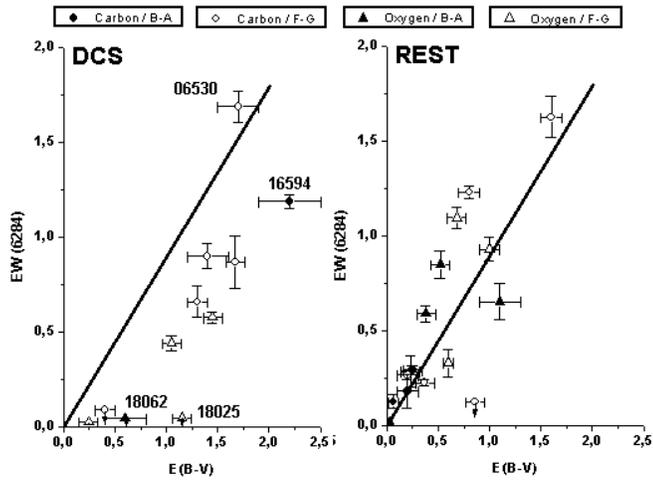}
\caption{
Equivalent width of the 6284~\AA\ band vs. $E(B-V)$ for the group of stars
dominated by circumstellar extinction (left panel; DCS type) and for the
rest of stars in the sample (right panel) with different symbols
indicating the dominant chemistry and spectral type of the observed stars.
The solid line represents the behaviour observed in field stars
dominated by interstellar extinction, as deduced from the data
shown in Figure \ref{fig:thor_02}.}
\label{fig:db_6284}
\end{center}
\end{figure}

For many cases DBs are observed towards the DCS-type stars
(Figs.~\ref{fig:db_6284} to~\ref{fig:db_6196}; left panels). In order to assess the
circumstellar contribution the observed $EW$ can be corrected by subtracting
the expected DIB $EW$ found by applying the estimated interstellar reddening
(Table~\ref{tb:extinction}) to the respective $EW$/$E(B-V)$ for field stars
(Table~\ref{tb:caracdbs}) as well as subtracting the IS reddening contribution
from the total observed reddening. The introduced uncertainties are quite large
due to the scatter on the derived linear relationships (see above,
Section~\ref{section:fieldstars}).
In particular \mbox{\object{IRAS 02229+6208}} and \mbox{\object{IRAS 08544-443}}
require a correction for significant IS dust.

In the other direction, we can correct the IS-type stars (left panel of
Figs.~\ref{fig:db_6284} to~\ref{fig:db_6196}) by subtracting the CS
(circumstellar) reddening
contribution from the total reddening and the circumstellar $EW$/$E(B-V)$
(which we assume to be zero) from the observed $EW$. It shows then that
all stars coincide neatly with the average Galactic relation.
This is most noticeable for \mbox{\object{IRAS 17086-403}} and \mbox{\object{IRAS 17395-0841}}
which have estimated CS contributions of 0.3 and 0.2~mag, respectively).

In principle, this result supports our initial interpretation that the DB
carrier(s) may not be present in the circumstellar envelopes of post-AGB stars.
However, strong variations from source to source are still visible in both
groups of stars, which may be related to other observational properties of the
shells not yet considered (Significant scatter is also observed for the
sample of field stars; Sect.~4.1).
Indeed, the results obtained suggest that some of the DB carriers could be completely
absent in some of these envelopes while not in others.

\subsection{DB strength vs. chemistry and spectral type}

In order to explore whether other environmental conditions, like the dominant
chemistry in the shell or the spectral type of the central star could also play
a role in the differences observed between individual stars in the sample, we
have further divided the two groups defined above in another four
subgroups as a function of whether the chemistry of the shell is carbon-rich or
oxygen-rich, or the spectra of the central star is of early-type
(B-A)\footnote{Note that among the subgroup of early-type stars we have also
included the few sources in Table~\ref{tb:referencias} which are classified as
planetary nebulae.} or of intermediate-type (F-G).

We do this because the dominant chemistry of the shell can completely determine
the formation of specific compounds in the circumstellar shell. In oxygen-rich
shells we expect to find aluminum oxides, amorphous or crystalline (fayalite,
enstatite, forsterite, etc) silicates, water ice and other main constituents of
oxygen-rich dust grains. In carbon-rich stars, instead, we can find
carbon-based constituents, like chains or rings of carbon, graphite,
hydrogenated amorphous carbon grains, fullerenes, nanodiamonds or PAHs.

If a DB carrier had their origin in a compound or constituent related to only
one of the above chemistries, we would expect to observe differences in
strength from source to source as a function of their particular chemical
composition.

On the other hand, it is also well known that the UV radiation field plays a
crucial role in the processing of the circumstellar dust grains, not only
immediately after they are formed, while they are still part of the shell, but
also later when they are released to the ISM. Dust grains in the circumstellar
envelopes of post-AGB stars are exposed to increasing doses of UV radiation due
to the increasing effective temperature of the central star during its fast
evolution towards the planetary nebula stage. First, when the central star is
still showing late to intermediate spectral type, the UV radiation can be
neglected, both as a consequence of the low effective temperature of the
central star and because the higher density in the envelope during the early
post-AGB stage would effectively protect (at least temporarily) circumstellar
dust grains from the energetic UV photons coming from the ISM. These conditions
favour the formation of large dust grains which can survive in this less
aggressive environment. Later in the post-AGB evolution the central stars
become early-type and they start producing a considerable number of UV photons
which may lead to an efficient processing of the dust grains in the shell,
which in turn is less protected and more vulnerable also to the UV radiation
field coming from the ISM. The combined effect of the UV photons coming from
the central star and from the ambient ISM is expected to accelerate the
processing of the dust grains, leading to new species like molecules, radicals
(more or less complex) and other byproducts resulting from the partial or total
evaporation of the grains, which will eventually be released to the ISM.
Indeed, these byproducts could be the actual carriers of the DIBs commonly
observed in the ISM.

If DB carriers are only related to the byproducts of the decomposition of these
large circumstellar grains, we would expect to observe a deficit in DB
strengths in post-AGB stars, only while the central stars are still showing a
relatively low effective temperature.

Both effects can be combined, and it may also happen that the DB carriers are
related to the byproducts of only a particular class of grains associated to a
given dominant chemistry. In this case, we will be able to detect significant
differences from source to source, both as a function of the spectral type of
the central star as well as of the dominant chemistry in the shell.

In the next section we will analyse the influence of these environmental
conditions (dominant chemistry and spectral type) on the observed results for
each of the 9 DBs under study in our sample of post-AGB stars.

\subsection{Analysis of individual bands}
\label{section:analisis}

\subsubsection{Analysis of the 6284 \AA\ band}

The 6284~\AA\ band is not only the strongest ($EW/E(B-V) = 0.90$~\AA/mag in the ISM)
but also the broadest band included in our analysis. As such, it is relatively
easy to measure, in spite of the contamination by telluric lines already shown
in Figure~\ref{fig:telluric}, which must be carefully removed.

In Figure~\ref{fig:db_6284} we show the results obtained as a function of the
dominant chemistry and of the spectral type of the central star for each of the
two main subgroups identified in our sample.

The strength of the 6284~\AA\ band as a function of $E(B-V)$ for the group of
stars dominated by circumstellar extinction is presented in the left panel,
while the results obtained for the rest of stars in our sample is shown in the
right panel. As we can see, it is obvious that the post-AGB stars belonging to
the DCS group show DB strengths systematically below those observed in the ISM
(represented by the solid line).

Actually, in some cases this band is so weak than we can only determine an
upper limit for its equivalent width. This is the case for
\mbox{\object{IRAS 17436+5003}} (oxygen-rich; F type), \mbox{\object{IRAS
18025-3906}} (oxygen-rich; G type) and \mbox{\object{IRAS 18062+2410}}
(oxygen-rich; B type). It must be noted that the quality of the available
spectra for these four stars is good enough to discard a non-detection which
could be attributed to a poor signal-to-noise ratio.

Even for \mbox{\object{IRAS 16594-4656}} (carbon-rich; B-type), the star with the
second strongest feature in our sample, we find that the 6284~\AA\ band is a
factor of two weaker than expected for its high value of $E(B-V)$. This star is
a well-known bipolar proto-planetary nebula which seems to be affected by a
high internal extinction.

For \mbox{\object{IRAS 06530-0213}} the band strength is typical for
the total observed reddening being due to interstellar dust. On the other hand,
the interstellar and circumstellar reddening contributions are estimated to be
1.3 and 0.4~mag, respectively (Sect.~\ref{sec:extinction}). Though this could indicate
the presence of circumstellar DBs it should be noted that this line of sight lies in
the galactic plane (GLAT = $-$0.13 degrees) and the interstellar reddening could
be underestimated by the extinction model.

Remarkably, we find stars with a very weak band in all the subgroups,
irrespective of the dominant chemistry and spectral type considered. This
almost completely rules out the possibility of the 6284~\AA\ band being
generated in the circumstellar envelope of post-AGB stars, at least in the same
proportion as in the ISM.

In contrast, in the right panel we can see that in general the rest of stars in
the sample show a trend which is very similar to the one observed in the field
stars in which the extinction is dominated by the interstellar contribution.

\subsubsection{Analysis of the 5780 \AA\ band}


\begin{figure}
\begin{center}
\includegraphics[width=\columnwidth]{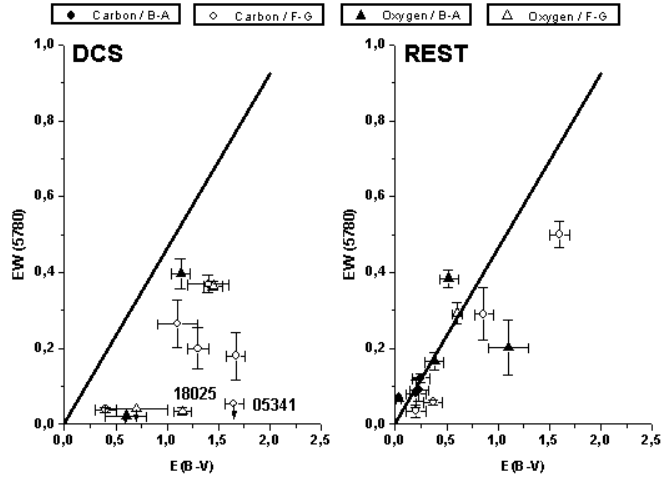}
\caption{Same as Figure \ref{fig:db_6284}, for the 5780~\AA\ band.}
\label{fig:db_5780}
\end{center}
\end{figure}

\begin{figure}
\begin{center}
\includegraphics[width=\columnwidth]{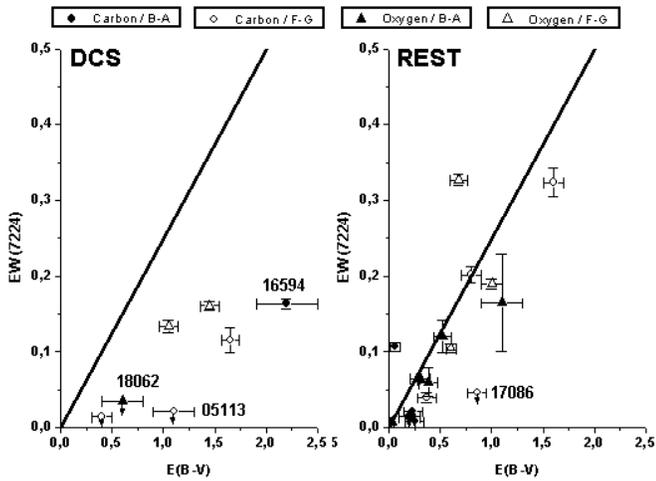}
\caption{Same as Figure \ref{fig:db_6284}, for the 7224~\AA\ band.}
\label{fig:db_7224}
\end{center}
\end{figure}

\begin{figure}
\begin{center}
\includegraphics[width=\columnwidth]{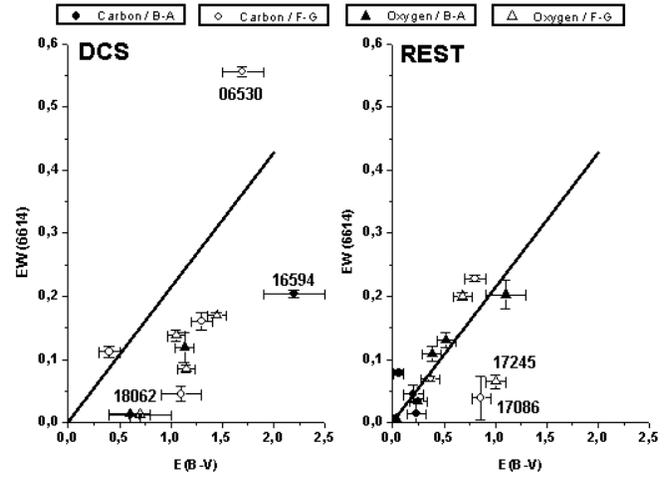}
\caption{Same as Figure \ref{fig:db_6284}, for the 6614~\AA\ band.}
\label{fig:db_6614}
\end{center}
\end{figure}

\begin{figure}
\begin{center}
\includegraphics[width=\columnwidth]{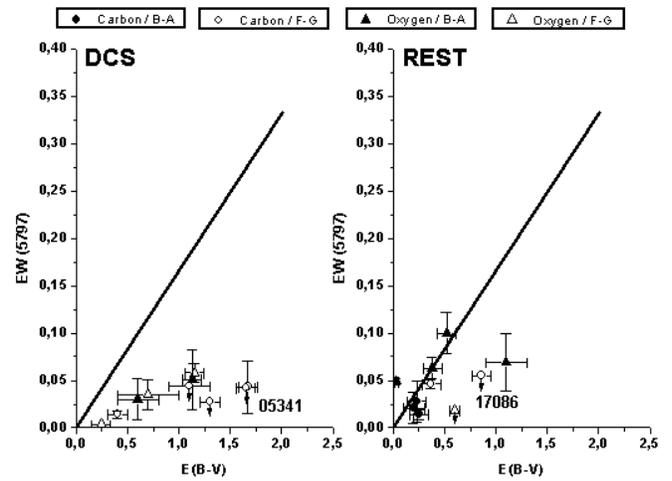}
\caption{Same as Figure \ref{fig:db_6284}, for the 5797~\AA\ band.}
\label{fig:db_5797}
\end{center}
\end{figure}

\begin{figure}
\begin{center}
\includegraphics[width=\columnwidth]{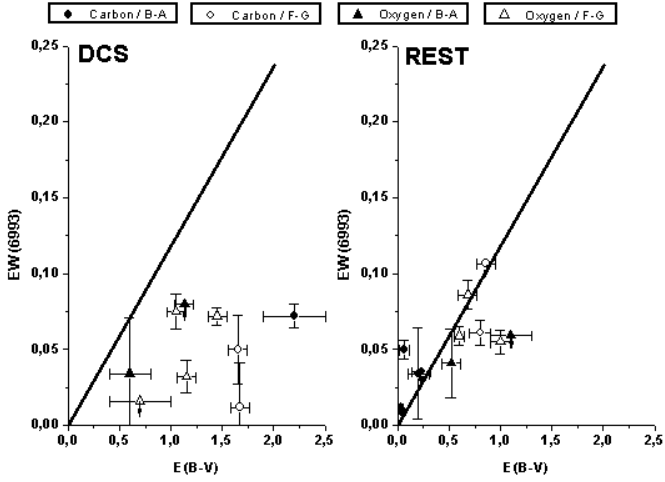}
\caption{Same as Figure \ref{fig:db_6284}, for the 6993~\AA\ band.}
\label{fig:db_6993}
\end{center}
\end{figure}

\begin{figure}
\begin{center}
\includegraphics[width=\columnwidth]{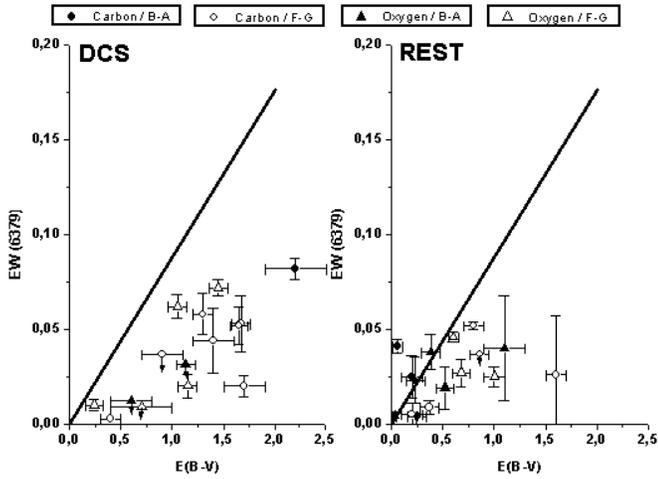}
\caption{Same as Figure \ref{fig:db_6284}, for the 6379~\AA\ band.}
\label{fig:db_6379}
\end{center}
\end{figure}

\begin{figure}
\begin{center}
\includegraphics[width=\columnwidth]{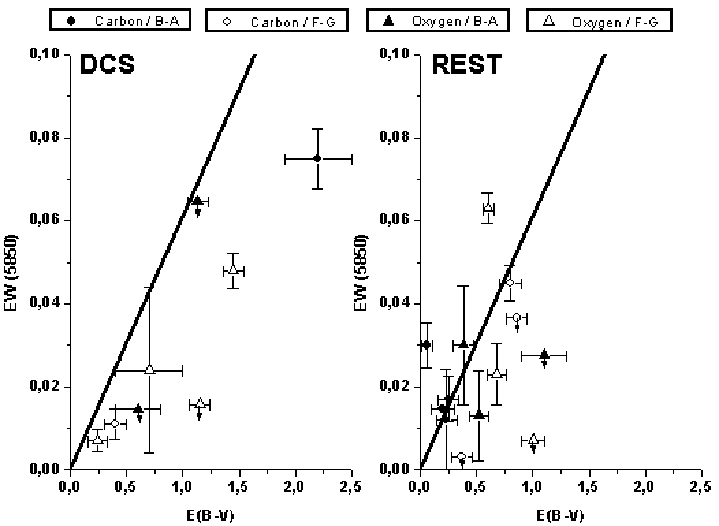}
\caption{Same as Figure \ref{fig:db_6284}, for the 5850 \AA~ band.}
\label{fig:db_5850}
\end{center}
\end{figure}

\begin{figure}
\begin{center}
\includegraphics[width=\columnwidth]{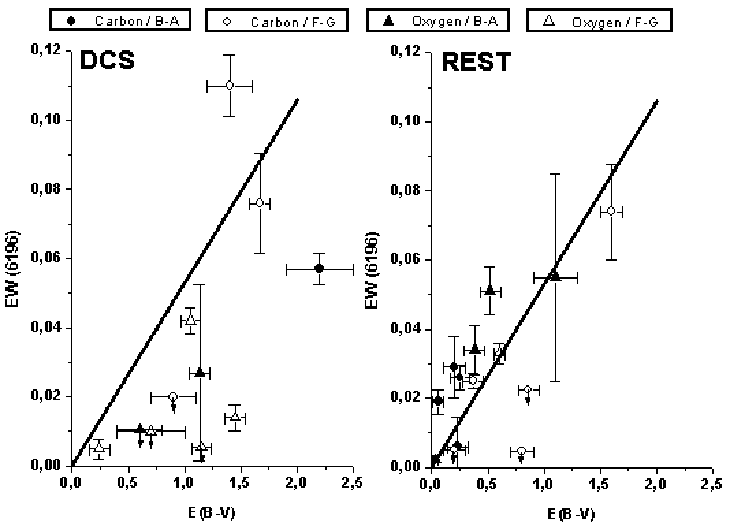}
\caption{Same as Figure \ref{fig:db_6284}, for the 6196 \AA~ band.}
\label{fig:db_6196}
\end{center}
\end{figure}


This DB has a $EW/E(B-V) = 0.46$~\AA/mag in the ISM, so it is the second most intense
after the 6284~\AA\ band. In this particular case, it is important to remark
that the spectral region corresponding to this band can be contaminated by the
presence of photospheric lines, which makes the evaluation of the band strength
very difficult in stars with intermediate and late spectral types.

Note that to distinguish weak features from weak stellar lines or telluric
contaminations is not always a simple task and makes it necessary to use
detailed stellar models (to subtract the atmospheric features) and high
resolution spectroscopy (to properly remove undesired contaminations), as the
only way to derive the accurate strength of the band, which is beyond the
scope of this work.

In Figure~\ref{fig:db_5780} we show the equivalent width of the 5780~\AA\ band
vs. $E(B-V)$ for each of the two main subgroups in which we have divided the
sample. In the group of stars dominated by circumstellar extinction (left
panel) we observe again strengths significantly weaker than those expected in
stars for which the extinction is mainly of interstellar origin, represented by
the solid line.

The non-detection of this DB in \mbox{\object{IRAS 05341+0852}} (carbon-rich;
F-type), despite the large value of $E(B-V)_{\rm CS}$ $\approx$~1.5~mag, is
remarkable. This clearly suggests that the carrier of this band is completely
absent at least in the envelopes of carbon-rich stars with intermediate
spectral types. A similar conclusion can be derived for oxygen rich stars with
intermediate spectral types from the very weak strength observed in
\mbox{\object{IRAS 18025-3906}} (oxygen-rich; B-type).

Unfortunately, the spectrum available for \mbox{\object{IRAS 16594-4656}} does
not cover the spectral range corresponding to this band, so we cannot extend
the above conclusion to carbon-rich stars with earlier spectral types based on
our data.

As in the case of the 6284~\AA\ band, we can also observe in the right panel of
Figure \ref{fig:db_5780} that the rest of stars in the sample not identified as
dominated by circumstellar extinction show a position in the diagram which is,
overall, in better agreement with the results obtained for field
stars in which the extinction is mainly of interstellar origin.

\subsubsection{Analysis of the 7224 \AA\ band}

This DB is not usually analysed in the literature because it is strongly
affected by telluric contamination. As in the case of the 6284~\AA\ band, we
have carefully eliminated this contribution by dividing the normalised spectrum
by the spectrum of the unreddened target \object{HD 172324}
(see Sect.~\ref{section:observation}).

In Figure~\ref{fig:db_7224} we show the equivalent width of the 7224~\AA\ band
vs. $E(B-V)$ for each of the two main subgroups identified in our sample. As
for the two previous features, we find strengths which are much weaker than
those measured in field stars in the subgroup formed by the stars in which the
circumstellar contribution to the overall extinction is dominant (DCS; left
panel). This again suggests that this band is not formed in the
circumstellar envelope of post-AGB stars.

Again, the measured intensity of the 7224~\AA\ band in \mbox{\object{IRAS
16594-4656}} is rather weak and, once more, we find several non-detections:
\mbox{\object{IRAS 05113+1347}} (carbon-rich star) and \mbox{\object{IRAS 18062+2410}}.
The latter case is
remarkable because this is an oxygen-rich star with a B spectral type, where we
have neither detected the 6284~\AA\ band. The non-detection of DBs in
oxygen-rich envelopes around post-AGB stars indicates that the DB carriers are
probably not generated in oxygen-rich environments. The absence of the band in
the specific case of \mbox{\object{IRAS 18062+2410}} suggests that they are
neither produced even when the central star temperature is hot enough to
produce high levels of UV irradiation on the (oxygen-rich) circumstellar
grains.

For the rest of stars in the sample (Figure~\ref{fig:db_7224}; right panel) we
find again a behaviour which seems to follow very well the general trend
observed in reference stars affected only by interstellar extinction.
The exception is \mbox{\object{IRAS 17086-2403}} (carbon rich star)
for which we detect no DBs.

\subsubsection{Analysis of the 6614 \AA\ band}

In Figure~\ref{fig:db_6614} we show the results of our analysis applied
this time to the 6614~\AA\ band.

In the left panel we show the equivalent widths measured in the subgroup of
post-AGB stars dominated by circumstellar extinction. As for the other bands,
we see strengths which are systematically weaker than in the reference stars
dominated by interstellar extinction, represented by the solid line in the
diagram. The only non-detection in this case corresponds to \mbox{\object{IRAS
18062+2410}}, which we have previously pointed out as non-detected in the
analysis made for the bands centred at 6284 and 7224~\AA. The weak strength of
this DB in \mbox{\object{IRAS 16594-4656}} (carbon-rich, B type) is found to be
again compatible with the absence of its carrier in the circumstellar envelope
of this star, and confirms that the overall extinction affecting this
source is the result of a quite similar contribution from the ISM and from the
circumstellar material.
The 6614~\AA\ DB detected toward \mbox{\object{IRAS 06530-0213}} is
unusually strong for DBs in the DCS group and  even with respect to the Galactic
relationship. As mentioned earlier for the 6284~\AA\ band toward the same target,
this could point, for this particular source, towards the presence of circumstellar
DBs or, perhaps more likely, an underestimation of the interstellar reddening.

For the rest of stars (right panel), as usual, we find that most of them are
located in the region of the diagram corresponding to the field stars dominated
by interstellar extinction.
In this case, we would like to remark only the slightly discrepant position
occupied by the oxygen-rich, F-type star \mbox{\object{IRAS 17245-3951}}
not yet previously identified as outlier in the above discussion.
Again, the carbon rich star \mbox{\object{IRAS 17086--2403}} shows
very weak DBs.

\subsubsection{Analysis of the 5797~\AA\ band}

This DB has been included in numerous studies in the literature because of its
proximity to the nearby 5780~\AA\ band. This has allowed a comparative analysis
of their relative intensities in different astrophysical environments.

The 5797~\AA\ band has a lower sensitivity to the extinction $EW/E(B-V) = 0.17$~\AA/mag
when it has been measured in the ISM, compared to the previous DBs. Similar
to the adjacent 5780~\AA\ band, it is necessary to take into account in our
analysis the possible contamination due to the presence of atmospheric stellar
lines in this spectral range in stars of intermediate and late spectral types,
as it can affect our measurements.

In Figure~\ref{fig:db_5797} we show the equivalent width of this band vs.
$E(B-V)$ as it has been measured from the available spectra for each subgroup
of stars in which we have divided the sample. For stars in the DCS group (left
panel), all post-AGB stars are found to show DB strengths which are
considerably weaker than in the field stars, consistently with the results
found in the other bands analysed so far.

Among the non-detections, we emphasize \mbox{\object{IRAS 05341+0852}}
(carbon-rich, F type) taking into account its large extinction
$E(B-V)_{\rm CS}$ = 1.5~mag. We recall that this star was also found to show no
indication of the presence of the accompanying feature at 5780~\AA.

Unfortunately, the spectrum of \mbox{\object{IRAS 16594-4656}} does not cover
the wavelength corresponding to this DB (as it happened with the 5780~\AA\ band).

In the right panel (rest sample) the line-of-sight toward \mbox{\object{IRAS 17086--2403}}
shows again a conspicuous absence of DBs.

\subsubsection{Analysis of the 6993~\AA\ band}

This band is also among the ones not usually analysed in the literature,
likely because of the presence of telluric lines in the spectral range adjacent
to this band but also because of the intrinsic weakness of this DB, for which
$EW/E(B-V) = 0.12$~\AA/mag in the ISM.

In Figure~\ref{fig:db_6993} we show the results of our analysis for this DB for
each of the subgroups in which we have divided the sample, again as a function
of the dominant chemistry and the spectral type of the central star.

The results obtained are once more consistent with previous analysis performed
for other DBs. We find a better agreement with the values obtained in reference
stars dominated by interstellar extinction for the sources in the right panel,
although in this case the effect is not so evident as in the previous analysis
due to the larger errors associated to the measurements.

Consistent results, although more sensitive to measurement errors, are obtained
when the features at 6379, 5850 and 6196~\AA\ are analysed
(see Figures~\ref{fig:db_6379}, \ref{fig:db_5850} and~\ref{fig:db_6196}).

\subsection{Radial velocity analysis}\label{section:velocidad}

\begin{figure}[t]
\begin{center}
\includegraphics[width=0.9\columnwidth]{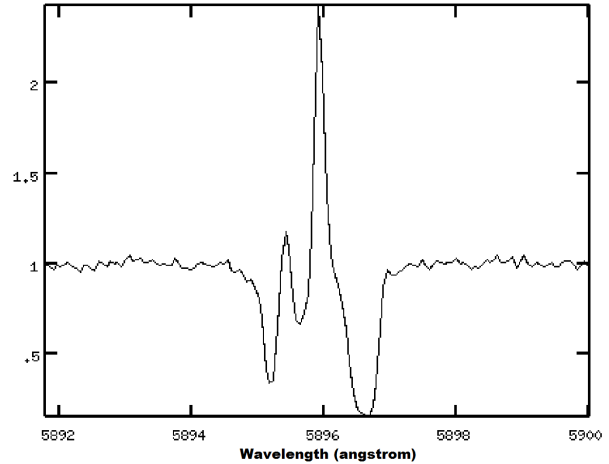}
\caption{The complex profile of the \ion{Na}{i}~D$_1$ line,
as observed in IRAS 04296+3429.}
\label{fig:sodio}
\end{center}
\end{figure}


\begin{figure}[t]
\centering
\includegraphics[bb=35 20 580 830,width=.8\columnwidth,clip]{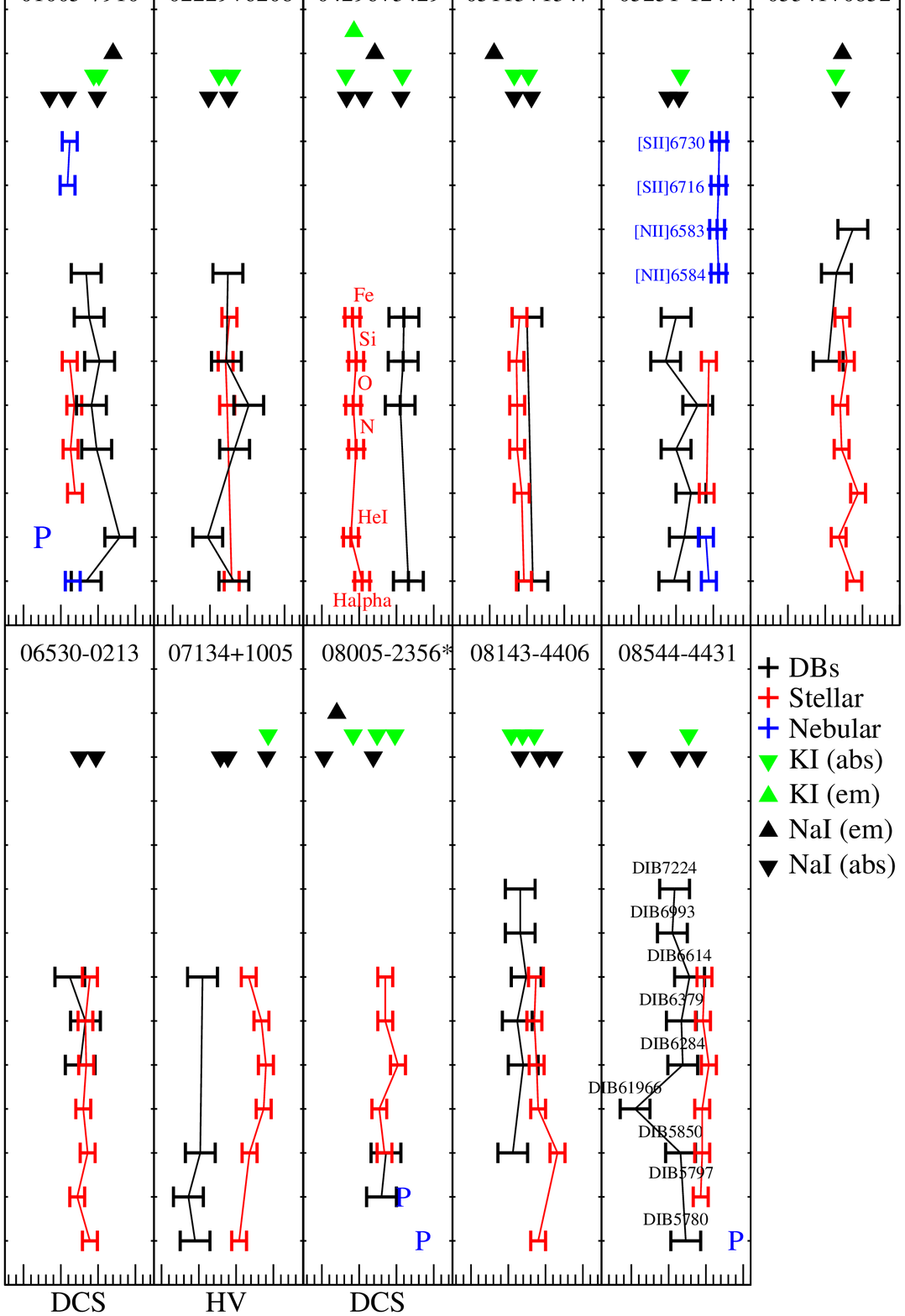}
\caption{
Velocities of the DBs and the stellar and interstellar absorption
lines and the nebular emission line components for each target are plotted per
panel. The y-axis is in arbitrary units. The x-axis is the LSR velocity in
km~s$^{-1}$, with big tick-marks separated by 50~km~s$^{-1}$. Note that
the width of the panels are identical, i.e. 200~km~s$^{-1}$, but the
central velocity of each is shifted to show all lines for each target.
Error bars are $\leq$10~km~s$^{-1}$ for stellar lines, $\leq$20~km~s$^{-1}$ for DBs
and $\leq$5~km~s$^{-1}$ for the sodium and potassium components.
Targets dominated by circumstellar reddening are
labeled `DCS' and high radial velocity targets are labeled `HV' The stellar and
DB velocity components can be directly compared to those of neutral sodium and
potassium in either emission (upward arrow) or absorption (downward arrow)
plotted at the top of each panel.}
\label{fig:velocities1}
\end{figure}

\addtocounter{figure}{-1}

\begin{figure}[t]
\centering
\includegraphics[bb=35 20 580 830,width=.8\columnwidth,clip]{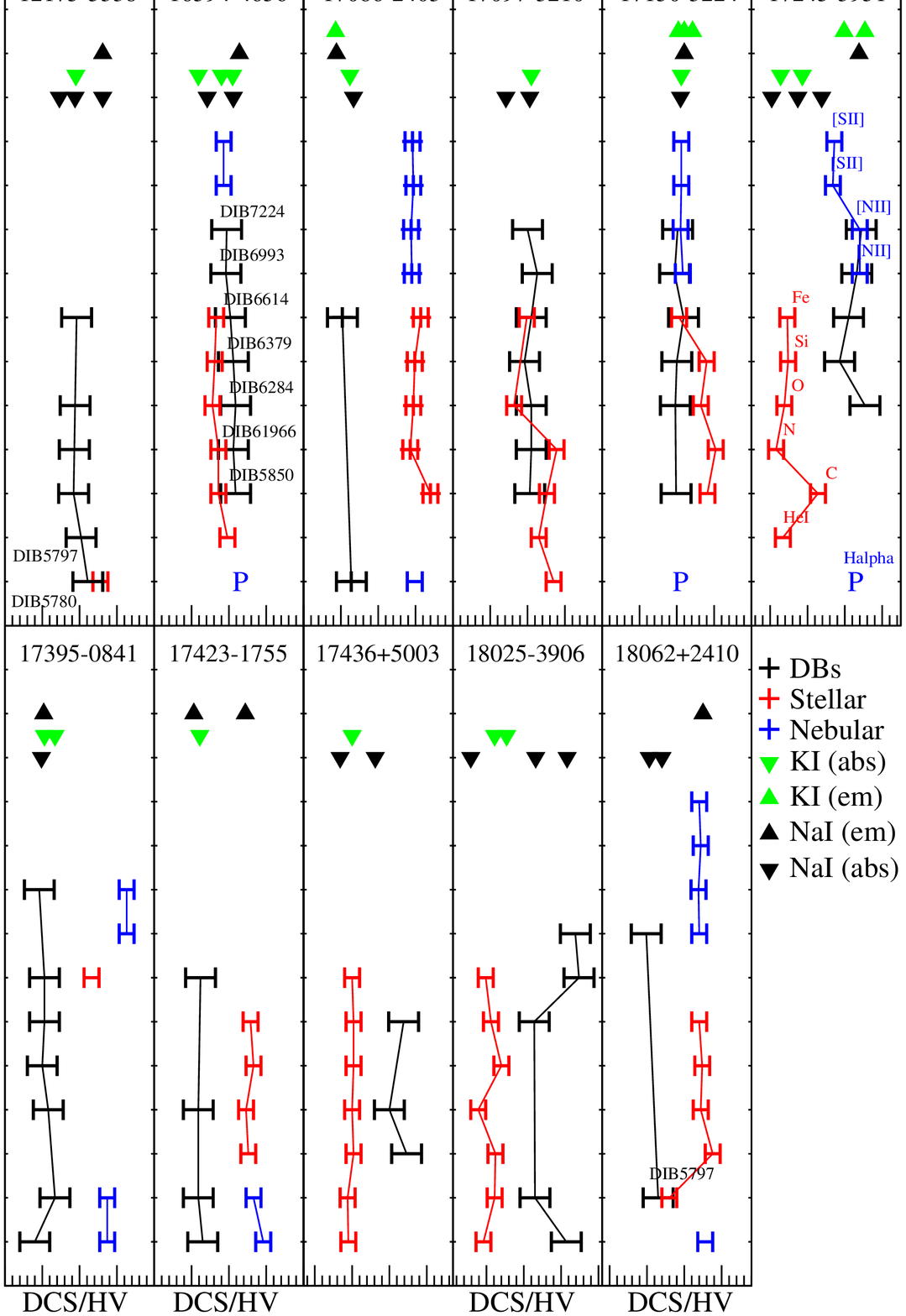}
\caption{(continued).}
\label{fig:velocities2}
\end{figure}

\addtocounter{figure}{-1}

\begin{figure}[t]
\centering
\includegraphics[bb=35 20 580 830,width=.8\columnwidth,clip]{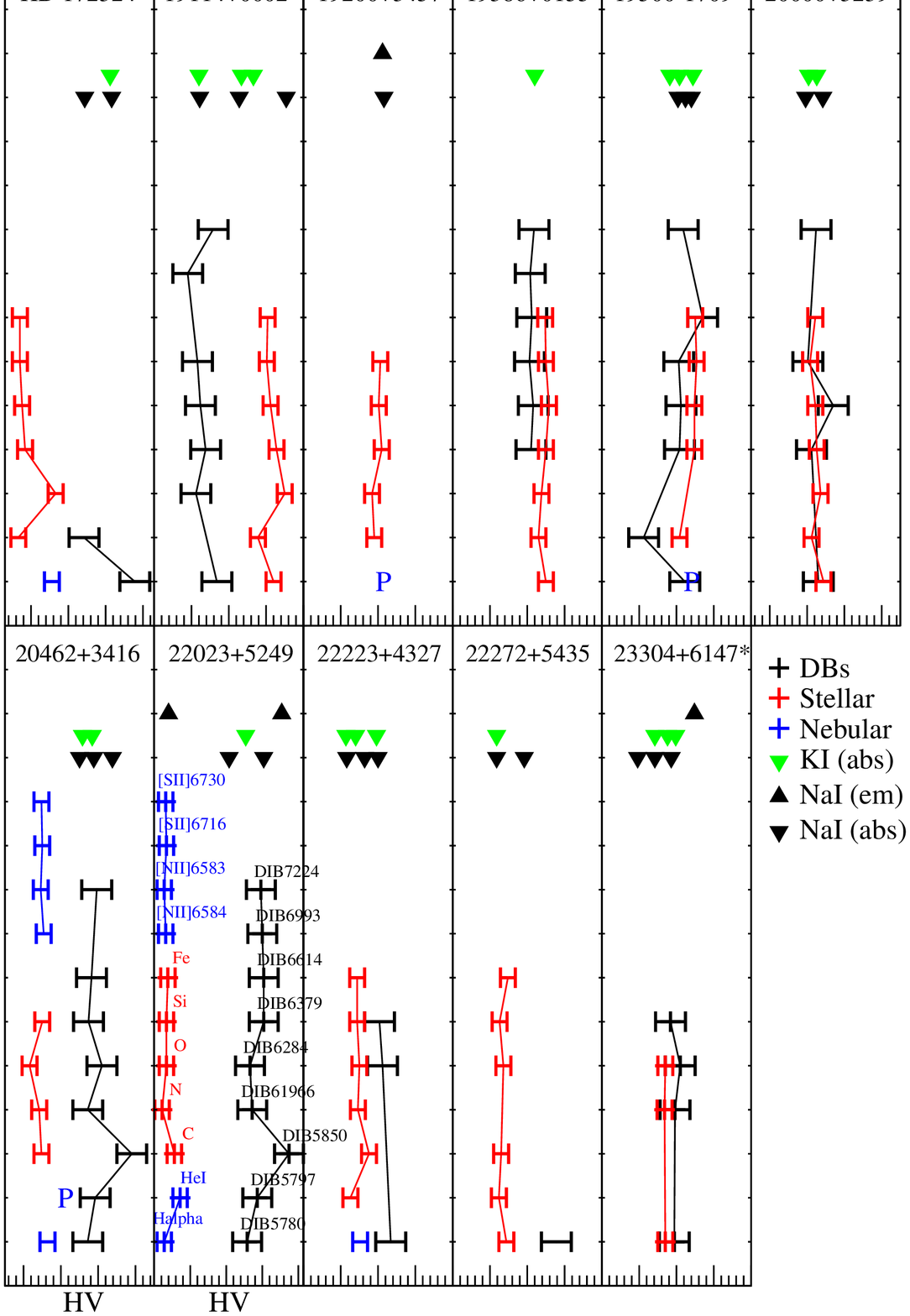}
\caption{(continued).}
\label{fig:velocities3}
\end{figure}

\begin{figure}[t]
\begin{center}
\includegraphics[width=0.8\columnwidth]{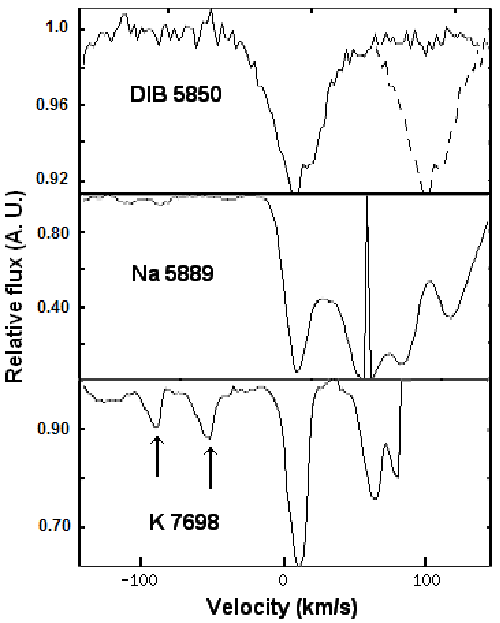}
\caption{
The 5850 $\AA$ band observed in the high radial velocity
($\sim$ 100 km~s$^{-1}$) sample star \mbox{\object{IRAS 19114+0002}}
(solid line).
The non-detection of this feature at the Doppler-shifted velocity
corresponding to the central star (dashed line) confirms the
interstellar nature of the band and the absence of a circumstellar
component.
The potassium and sodium velocity structures are
shown in the bottom panels in order to illustrate the
distribution of interstellar medium in this line of sight.
A narrow sodium emission line is seen in the middle panel.
The two arrows in the bottom most panel indicate two telluric features
next to the potassium line.}
\label{fig:velocidad}
\end{center}
\end{figure}


An additional way to check whether our conclusions are consistent with the
available observational data is to analyse the Doppler velocities associated to
the DBs detected in our stars.

The overall strategy consists of comparing these Doppler velocities with the
radial velocities associated either to the atmospheric stellar absorptions or
to the nebular and recombination emission lines sometimes detected in our
spectra. In general, atmospheric and nebular lines are expected to match each
other within the errors unless the central star is part of a binary system or
the nebular shell shows a complex morphology.

If the DBs detected are formed in the circumstellar envelopes of these stars,
we should measure Doppler velocities in these bands consistent with the
characteristic radial velocities derived from the absorption and/or emission
lines identified in the stellar spectra.

In Table~\ref{tb:velest1} (Online only) we give the radial velocities (in km~s$^{-1}$),
measured with respect to the Local Standard of Rest (LSR),
associated to several atoms and ions, as derived from various atmospheric
stellar absorptions and nebular emission lines identified in the stars of our
sample. In addition, we also display the measurements made in $H\alpha$ at
6563~\AA\ and in the \ion{He}{i} line at 6678~\AA. In the case of the
atmospheric absorption lines shown in Table \ref{tb:velest1} the average
velocity derived from several line measurements corresponding to various ions
of the elements C, N, O, Si and Fe is presented. For the nebular lines, we have
only considered the forbidden lines of [\ion{N}{ii}] and [\ion{S}{ii}], found
around $H\alpha$. The typical uncertainties are of the order of 5$-$10
km~s$^{-1}$.

In addition, in Table~\ref{tb:velest2} (Online only) we present the velocities
derived from the analysis of the \ion{Na}{i}\,D (5889.95 and 5895.92~\AA)
doublet and of the \ion{K}{i} (7698.97~\AA) line, which are in most cases also
well detected in our spectra (uncertainties are $\sim$~3 to 5~km~s$^{-1}$).
These lines, like the DBs that we want to analyse, usually originate in the
ISM, but they can also form in the circumstellar shell. In this case, the
circumstellar component usually appears in emission over the interstellar
absorption (see Figure \ref{fig:sodio}). In general, these lines show very
complex absorption profiles as they reflect the different velocities of the
clouds located along the line of sight. In some of our stars the circumstellar
component may contribute significantly to the observed profile and can be used
as a further test to identify the origin of analogue velocity components which
may be present in our favourite DB.

Table~\ref{tb:veldib} (Online only) shows the
radial velocities associated to
the DBs observed in the stars of our sample, which can then be compared to the
velocities provided in Tables \ref{tb:velest1} and \ref{tb:velest2}.

It is important to take into account that deriving velocities for DBs
is in many cases a complicated task, especially
if the features under analysis are weak in strength. In general, the Doppler
shift measurements are determined by assuming that the absorption peak is a
good approximation to the centre of the feature. We estimate that, on average,
the errors in Table~\ref{tb:veldib} may be affected by errors of the order of
10$-$20~km~s$^{-1}$.

Comparing Tables~\ref{tb:velest1} and~\ref{tb:velest2} we observe that in most
cases there is at least one velocity component associated to the sodium doublet
or the potassium line, either in emission or in absorption, that can be
interpreted as having a stellar or circumstellar origin. The circumstellar
nature of these lines is easy to determine when they are found in emission. The
radial velocities measured in this case are usually coincident with the
systemic velocity of the post-AGB star. The few cases found in which our
measurements do not support this statement are indicated with an asterisk in
Table~\ref{tb:velest2} and Fig.~\ref{fig:velocities1}.
They correspond to very complex \ion{Na}{i}\,D line
profiles in which the circumstellar emission appears over-imposed to the
interstellar absorption.

Figure~\ref{fig:velocities1} shows, for ease of comparison, the velocities
of the DBs and the (inter-)stellar absorption and emission components for each
target. For the majority of the targets these graphs show consistent velocities
for the stellar lines. DB velocities are also consistent with each other. For
several cases the nebular (emission) lines are significantly shifted with
respect to the atmospheric lines (\emph{e.g.} \mbox{\object{IRAS 17245-3951}})
due to binarity of the system and/or a complex wind structure.
These stellar and DB velocity components can be directly compared to those of
neutral sodium and potassium in the respective line-of-sight.

If DB carriers are present in the circumstellar envelopes of some of the
post-AGB stars in our sample, we would expect to find as well matches between
the velocities shown in Table~\ref{tb:veldib} and those in
Table~\ref{tb:velest1} (see Figure \ref{fig:velocidad}), especially for those
stars in which we have detected circumstellar \ion{Na}{i} in emission belonging
to the DCS group. Remarkably, in not any case we find values consistent with
the velocities associated to the DBs which cannot be explained as a natural
consequence of interstellar clouds with a similar velocity present in the line
of sight.

The inconsistency between velocities is more obvious if we have a look at those
stars showing very high radial velocities (HV in Table~\ref{tb:extinction}
and Fig.~\ref{fig:velocities1}).
Several of the HV targets have radial velocities larger than 100~km~s$^{-1}$.
Such large velocity differences are comparable to those measured
for successfully detected extra-galactic DBs (\citealt{Ehrenfreund02,Cox07}).

In the top panel of Fig.~\ref{fig:velocidad} we illustrate the radial
velocities difference expected between the interstellar (solid line) and
circumstellar (dashed line) DBs for the HV target \mbox{\object{IRAS 19114+0002}}.
The narrow 5850~\AA\ CS DB (shifted to the stellar radial velocity) would be well
separated from the observed (IS) DB. The atomic line profiles of sodium (middle
panel) and potassium (bottom panel) are shown to indicate the ISM distribution
in this line of sight. Note also the narrow sodium emission.

In none of the stars is it possible to assign DBs to nebular or
stellar lines exclusively. This strongly supports our conclusion that the
DBs detected toward the post-AGB stars in our sample are not originated in their
circumstellar envelopes.

Globally considered, the radial velocity analysis here presented gives more
strength to our proposed scenario, in which the DB carriers are suggested to be
not present in the circumstellar envelopes of post-AGB stars, or at least not
under the excitation conditions necessary to produce the transitions that we
identify as DBs in the ISM. Targets that show large velocity differences
between interstellar and circumstellar lines and that show significant
circumstellar reddening (\emph{e.g.} \mbox{\object{IRAS 17086-2403}},
\mbox{\object{IRAS 17423-1755}}, \mbox{\object{IRAS 18025-3906}} and
\mbox{\object{IRAS 18062+2410}}) provide the best candidates to search for the
presence of (weak) circumstellar DBs separated from the interstellar DBs. Our
current spectra are of insufficient quality to search for these weak features
next to the observed DIBs. Note that both CS and IS DBs could coexist.
And, if separated by more than their FWHM ($\sim 40 - 60$~km~s$^{-1}$ for
narrow DBs) the central velocity of the IS and (possibly) CS DB would not be
affected by each other.

\section{Conclusions}\label{section:conclusions}

The equivalent widths of 9 DBs commonly found in the ISM have been determined
for a representative sample of galactic post-AGB stars displaying a wide
variety of observational properties. We present here the results of
our extensive survey to look for DBs in envelopes of evolved stars.

We have carefully disentangled the observed extinction by assessing the
expected interstellar extinction for each of the observed targets. This allowed
us to select a sub-sample of targets whose line of sight reddenings are
dominated ($>50\%$) by circumstellar dust.

In general, the strengths of the DBs are found to follow the same correlation
with $E(B-V)$ observed in field stars only in those sources showing little
circumstellar contribution to the overall reddening. In contrast, DBs are weak
or absent in sources dominated by circumstellar reddening, irrespective 
of the dominant chemistry and spectral type of the central star,
although our conclusions should be taken with caution due to
the relatively small sample size.

The results obtained suggest that the carrier(s) of the DBs do not form or at
least they are not ``available'' to produce any detectable spectral feature
during the post-AGB phase. The carriers, if present in the circumstellar
envelope of these stars are not found under the environmental conditions needed
to excite the transitions which we identify as DBs in the ISM.

The radial velocity analysis of the features observed in individual
sources confirm this result, as the Doppler shifts measured are always found to
be consistent with an interstellar origin for the bands observed.

DB carriers may be carbonaceous species or radicals attached to large organic
molecules, trapped in lattice or more complex structures, or constituents of
the mantle of circumstellar dust grains which are liberated to the ISM only
after strong UV irradiation (either UV photons from the central star or from
the more energetic interstellar UV field).

In this sense, the identification of the carriers as strongly ionised PAHs
and/or radicals liberated from carbonaceous species as a consequence of
photo-evaporation of dust grains in the ISM looks tempting and would be
consistent with our observations.

However, we do not find any evidence of the carbonaceous nature of the
carrier(s) in our sample stars, something generally accepted in the literature,
nor any correlation with the presence of PAHs in the mid-infrared spectrum of
these sources, as it has been claimed by several authors in the past.

If DBs are connected with PAHs or with any other carbonaceous species such as
the ones suggested in the introduction of this paper, their carrier(s) must
form at a later stage, probably under different excitation conditions, once the
envelope of the post-AGB star is totally diluted in the interstellar medium as
a result of the expansion of the shell.


\begin{acknowledgements}
Many of the spectra used in the analysis here presented were kindly provided by
Hans van Winckel and Maarten Reyniers, working at the Katholieke Universiteit
Leuven, Belgium. The authors are also grateful to Bernard Foing and Nathalie
Boudin, who participated in the early stage of this project and with whom we
had very fruitful discussions.
We sincerely thank the referees for their helpful and constructive comments.
This work was partially funded by grants AYA2003--09499 and AYA2004--05382 of 
the Spanish Ministerio de Ciencia y Tecnolog\'\i a.
\end{acknowledgements}

\bibliographystyle{aa}

\Online

\begin{sidewaystable*}
\caption{Equivalent width measurements (in \AA) corresponding to the 9 DBs analysed
in our sample of post-AGB stars.}
\label{tb:intensidades}
\centering
\begin{tabular}{l r r r r r r r r r}\hline\hline 
\multicolumn{1}{c}{}    & \multicolumn{9}{c}{$EW$ (\AA)}\\
IRAS Name     &   5780~~~~        &   5797~~~~~         &   5850~~~~~            &   6196~~~~~            &   6284~~~~            &   6379~~~~            &   6614~~~~            &   6993~~~~            &   7224~~~~            \\
\hline
 01005$+$7910   &   0.08    $\pm$   0.04    &   0.021   $\pm$   0.009   &   $\le$       0.01    &   0.029   $\pm$   0.009   &   0.18    $\pm$   0.09    &   0.025   $\pm$   0.009   &   0.045   $\pm$   0.009   &   0.034   $\pm$   0.009   &   $\le$       0.01    \\
 02229$+$6208   &   0.18    $\pm$   0.06    &   0.043   $\pm$   0.009   &   ?           &   0.076   $\pm$   0.009   &   0.87    $\pm$   0.09    &   0.053   $\pm$   0.009   &   ?           &   0.012   $\pm$   0.009   &   ?           \\
 04296$+$3429   &   0.20    $\pm$   0.06    &   $\le$       0.03    &   ---         &   ?           &   0.66    $\pm$   0.08    &   0.058   $\pm$   0.009   &   0.16    $\pm$   0.01    &   ---         &   ?           \\
 05113$+$1347   &   0.27    $\pm$   0.06    &   $\le$       0.04    &   ?           &   ?           &   ?           &   ?           &   0.045   $\pm$   0.009   &   ---         &   $\le$       0.02    \\
 05251$-$1244   &   0.09    $\pm$   0.03    &   0.028   $\pm$   0.009   &   0.012   $\pm$   0.009   &   0.006   $\pm$   0.009   &   0.28    $\pm$   0.08    &   0.023   $\pm$   0.009   &   0.014   $\pm$   0.009   &   $\le$       0.04    &   $\le$       0.02    \\
 05341$+$0852   &   $\le$       0.06    &   $\le$       0.04    &   ---         &   ?           &   ?           &   0.052   $\pm$   0.009   &   ?           &   0.050   $\pm$   0.009   &   0.12    $\pm$   0.02    \\
 06530$-$0213   &   ---         &   ---         &   ?           &   ?           &   1.69    $\pm$   0.08    &   0.020   $\pm$   0.006   &   0.556   $\pm$   0.008   &   ---         &   ---         \\
 07134$+$1005   &   0.039   $\pm$   0.006   &   0.014   $\pm$   0.004   &   0.011   $\pm$   0.004   &   ?           &   $\le$       0.09    &   $\le$       0.003   &   0.112   $\pm$   0.009   &   ---         &   $\le$       0.01    \\
 08005$-$2356   &   $\le$       0.04    &   0.035   $\pm$   0.009   &   0.024   $\pm$   0.009   &   $\le$       0.01    &   ?           &   $\le$       0.009   &   $\le$       0.01    &   $\le$       0.02    &   ?           \\
 08143$-$4406   &   ---         &   ---         &   0.045   $\pm$   0.004   &   $\le$       0.005   &   1.23    $\pm$   0.03    &   0.052   $\pm$   0.002   &   0.228   $\pm$   0.005   &   0.061   $\pm$   0.008   &   0.202   $\pm$   0.009   \\
 08544$-$4431   &   0.37    $\pm$   0.01    &   ?           &   0.048   $\pm$   0.004   &   0.014   $\pm$   0.004   &   0.58    $\pm$   0.03    &   0.072   $\pm$   0.004   &   0.170   $\pm$   0.003   &   0.072   $\pm$   0.006   &   0.161   $\pm$   0.007   \\
 12175$-$5338   &   0.12    $\pm$   0.01    &   0.014   $\pm$   0.006   &   0.017   $\pm$   0.005   &   0.026   $\pm$   0.004   &   0.29    $\pm$   0.02    &   $\le$       0.004   &   0.034   $\pm$   0.004   &   ---         &   $\le$       0.008   \\
 16594$-$4656   &   ---         &   ---         &   0.075   $\pm$   0.007   &   0.057   $\pm$   0.005   &   1.19    $\pm$   0.04    &   0.082   $\pm$   0.006   &   0.204   $\pm$   0.007   &   0.072   $\pm$   0.008   &   0.164   $\pm$   0.007   \\
 17086$-$2403   &   0.29    $\pm$   0.07    &   $\le$   $\pm$   0.06    &   $\le$       0.04    &   $\le$       0.02    &   $\le$       0.1 &   $\le$       0.04    &   0.04    $\pm$   0.03    &   $\le$       0.1 &   $\le$       0.05    \\
 17097$-$3210   &   ---         &   ---         &   0.030   $\pm$   0.006   &   0.019   $\pm$   0.004   &   0.13    $\pm$   0.04    &   0.041   $\pm$   0.004   &   0.079   $\pm$   0.005   &   0.050   $\pm$   0.006   &   0.107   $\pm$   0.006   \\
 17150$-$3224   &   ---         &   ---         &   0.023   $\pm$   0.007   &   ?           &   1.10    $\pm$   0.06    &   0.027   $\pm$   0.007   &   0.200   $\pm$   0.006   &   0.086   $\pm$   0.009   &   0.327   $\pm$   0.007   \\
 17245$-$3951   &   ---         &   ---         &   $\le$       0.007   &   ?           &   0.93    $\pm$   0.06    &   0.025   $\pm$   0.006   &   0.065   $\pm$   0.009   &   0.055   $\pm$   0.008   &   0.190   $\pm$   0.007   \\
 17395$-$0841   &   0.20    $\pm$   0.07    &   0.07    $\pm$   0.03    &   $\le$       0.03    &   0.06    $\pm$   0.03    &   0.65    $\pm$   0.09    &   0.040   $\pm$   0.009   &   0.20    $\pm$   0.02    &   $\le$       0.06    &   0.165   $\pm$   0.065   \\
 17423$-$1755   &   0.40    $\pm$   0.04    &   0.05    $\pm$   0.03    &   $\le$       0.06    &   0.027   $\pm$   0.009   &   ?           &   $\le$       0.03    &   0.12    $\pm$   0.02    &   $\le$       0.08    &   ?           \\
 17436$+$5003   &   ?           &   $\le$       0.003   &   0.007   $\pm$   0.003   &   0.005   $\pm$   0.003   &   $\le$       0.02    &   0.010   $\pm$   0.003   &   ?           &   ---         &   ?           \\
 18025$-$3906   &   0.034   $\pm$   0.009   &   0.058   $\pm$   0.009   &   $\le$       0.02    &   $\le$       0.005   &   $\le$       0.05    &   0.020   $\pm$   0.006   &   0.085   $\pm$   0.007   &   0.032   $\pm$   0.009   &   ?           \\
 18062$+$2410   &   ?       0.02    &   0.03    $\pm$   0.02    &   $\le$       0.01    &   $\le$       0.01    &   $\le$       0.04    &   $\le$       0.01    &   $\le$       0.01    &   0.034   $\pm$   0.009   &   $\le$       0.04    \\
   HD 172324    &   0.068   $\pm$   0.004   &   0.049   $\pm$   0.004   &   ?           &   $\le$       0.002   &   $\le$       0.02    &   $\le$       0.004   &   $\le$       0.006   &   $\le$       0.01    &   $\le$       0.007   \\
 19114$+$0002   &   0.29    $\pm$   0.03    &   $\le$       0.02    &   0.063   $\pm$   0.004   &   0.033   $\pm$   0.003   &   0.33    $\pm$   0.07    &   0.046   $\pm$   0.002   &   ?           &   0.059   $\pm$   0.006   &   0.105   $\pm$   0.006   \\
 19200$+$3457   &   ?           &   ?           &   ?           &   ?           &   ?           &   ?           &   ?           &   ?           &   $\le$       0.07    \\
 19386$+$0155   &   ---         &   ---         &   ---         &   0.042   $\pm$   0.004   &   0.44    $\pm$   0.04    &   0.062   $\pm$   0.006   &   0.138   $\pm$   0.009   &   0.075   $\pm$   0.009   &   0.134   $\pm$   0.008   \\
 19500$-$1709   &   0.058   $\pm$   0.006   &   0.046   $\pm$   0.004   &   $\le$       0.003   &   0.025   $\pm$   0.002   &   0.23    $\pm$   0.02    &   0.009   $\pm$   0.004   &   0.069   $\pm$   0.004   &   ---         &   0.040   $\pm$   0.007   \\
 20000$+$3239   &   0.50    $\pm$   0.03    &   ?           &   ?           &   0.074   $\pm$   0.009   &   1.63    $\pm$   0.09    &   0.026   $\pm$   0.009   &   ?           &   ---         &   0.32    $\pm$   0.02    \\
 20462$+$3416   &   0.16    $\pm$   0.02    &   0.063   $\pm$   0.009   &   0.030   $\pm$   0.009   &   0.034   $\pm$   0.007   &   0.59    $\pm$   0.04    &   0.038   $\pm$   0.009   &   0.109   $\pm$   0.009   &   ---         &   0.06    $\pm$   0.02    \\
 22023$+$5249   &   0.38    $\pm$   0.02    &   0.10    $\pm$   0.02    &   0.013   $\pm$   0.009   &   0.051   $\pm$   0.007   &   0.85    $\pm$   0.07    &   0.019   $\pm$   0.009   &   0.130   $\pm$   0.009   &   0.041   $\pm$   0.009   &   0.12    $\pm$   0.02    \\
 22223$+$4327   &   0.035   $\pm$   0.009   &   ?           &   ?           &   $\le$       0.005   &   0.27    $\pm$   0.03    &   0.005   $\pm$   0.006   &   ?           &   ---         &   ?           \\
 22272$+$5435   &   ?           &   ?           &   ?           &   $\le$       0.02    &   ?           &   $\le$       0.04    &   ?           &   ---         &   ?           \\
 23304$+$6147   &   0.37    $\pm$   0.02    &   ?           &   ?           &   0.110   $\pm$   0.009   &   0.90    $\pm$   0.07    &   0.044   $\pm$   0.009   &   ?           &   ---         &   ?           \\
\hline
\multicolumn{10}{l}{ {\scriptsize ?: contamination by atmospheric stellar lines or poor S/N; }  {\scriptsize ---:  spectral range not covered}.}\\
\end{tabular}
\end{sidewaystable*}

\begin{table*}
\caption{Radial velocity measurements (in km s$^{-1}$)
associated to several atmospheric stellar and
nebular lines detected in the post-AGB stars of our sample.}
\label{tb:velest1}
\centering
\begin{tabular}{lrrrrrrrrrrr}\hline\hline
    &   H$\alpha$~~       &  \ion{He}{i}      &   C   &   N   &   O   &   Si  &   Fe  &  [\ion{N}{ii}]  &  [\ion{N}{ii}]  &   [\ion{S}{ii}]  &   [\ion{S}{ii}]  \\
IRAS Name &   6563        &   6678        &       &       &       &       &       &   6548    &   6583    &   6716    &   6730    \\
\hline
 01005$+$7910 &   $-$34 e   &   $-$76  P &   $-$31 &   $-$37 &   $-$32 &   $-$38 &   ---     &   ---  &   ---  &   $-$41 &   $-$38 \\
 02229$+$6208 &   29  a   &      ---      &   ---     &  ---      &   23  &   21  &   26  &   ---  &   ---  &   ---  &   ---  \\
 04296$+$3429 &   4 a   &   $-$11 a   &   ---     &   $-$4   &   $-$8    &   $-$4   &   $-$9    &   ---  &   ---  &   ---  &   ---  \\
 05113$+$1347 &   $-$4    a   &       ---   &   $-$7    &   $-$13 &   $-$13 &   $-$14 &   $-$10    &   ---  &   ---  &   ---  &   ---  \\
 05251$-$1244 &   44  e   &   40  e   &   41  &   ---     &   ---     &   44  &   ---     &   32,  57 &   31,  55 &   31,  57 &   32,  58 \\
 05341$+$0852 &   39  a   &   18  a   &   44  &   22  &   20  &   29  &   23  &   ---  &   ---  &   ---  &   ---  \\
 06530$-$0213 &   39  a   &   22  a   &   36  &   30  &   34  &   33  &   39  &   ---  &   ---  &   ---  &   ---  \\
 07134$+$1005 &   54  a   &       ---  &   68  &   87  &   90  &   84  &   67  &   ---  &   ---  &   ---  &   ---  \\
 08005$-$2356 &   85  P  &   59    P &   34  &   27  &   52  &   35  &   35  &   ---  &   ---  &  ---  & ---  \\
 08143$-$4406 &   40  a   &       ---  &   66  &   40  &   38  &   35  &   37  &   ---  &   ---  &   ---  &   ---  \\
 08544$-$4431 &   90 P &   43  a   &   45  &   45  &   54  &   46  &   48  &   ---  &   ---  &   ---  &   ---  \\
 12175$-$5338 &   28  a   &       ---   &   ---     &    ---    &    ---    &   ---     &    ---    &   ---  &   ---  &   ---  &   ---  \\
 16594$-$4656 &   15  P &   $-$2  a   &   $-$14 &   $-$14 &   $-$22 &   $-$19 &   $-$17 &   ---  &   ---  &   $-$7  &   $-$7  \\
 17086$-$2403 &   99  e   &       ---  &   120 &   93  &   97  &   99  &   107 &   95  &   94  &   97  &   96  \\
 17097$-$3210 &   35  a   &   15  a   &   26  &   39  &   $-$18 &    ---    &  $-$1  &   ---  &   ---  &   ---  &   ---  \\
 17150$-$3224$^*$ &   5   P &       ---  &   41  &   52  &   32  &   40  &   3   &   8   &   5   &   6   &   6   \\
 17245$-$3951$^*$ &   14  P &   $-$83 a   &   $-$36 &   $-$92 &   $-$81 &   $-$76 &   $-$77 &   20  &   20  &   $-$16 &   $-$14 \\
 17395$-$0841 &   87  e   &   87  e   &   ---     &    ---    &   ---     &    ---    &   66  &   113 &   113 &   ---  &   ---  \\
 17423$-$1755 &   96  e   &   83  e   &   76  &   73  &   83  &   79  &   ---     &   ---  &   ---  &   ---  &   ---  \\
 17436$+$5003 &  $-$55 a   &   $-$56 a   &   $-$48 &   $-$50 &   $-$48 &   $-$48 &   $-$50 &   ---  &   ---  &   ---  &   ---  \\
 18025$-$3906 &   $-$104    a   &   $-$89 a   &   $-$88 &   $-$111    &   $-$80 &   $-$94 &   $-$101    &   ---  &   ---  &   ---  &   ---  \\
 18062$+$2410 &   78  e   &   30  a   &   88  &   72  &   74  &   70  &   ---     &   70  &   69  &   72  &   70  \\
 HD 172324    &   $-$72 e   &   $-$117    a   &   $-$67 &   $-$108    &   $-$112    &   $-$115    &   $-$115    &   ---  &   ---  &   ---  &   ---  \\
 19114$+$0002 &   110 a   &   89  a   &   125 &   114 &   106 &   101 &   102 &   ---  &   ---  &   ---  &   ---  \\
 19200$+$3457 &   7   P &   $-$5  a   &   $-$8  &   5   &   1   &   3   &  ---      &   ---  &   ---  &   ---  &   ---  \\
 19386$+$0155 &   25  a   &   15  a   &   19  &   25  &   29  &   25  &   24  &   ---  &   ---  &   ---  &   ---  \\
 19500$-$1709 &   20  P &   4   a   &   ---     &   24  &   24  &   27  &   25  &   ---  &   ---  &   ---  &   ---  \\
 20000$+$3239 &   22  a   &   6   a   &   18  &   13  &   11  &   4   &   11  &   ---  &   ---  &   ---  &   ---  \\
 20462$+$3416 &   $-$68 e   &   $-$44  P &   $-$76 &   $-$79 &   $-$92 &   $-$75 &   ---     &   $-$73 &   $-$77 &   $-$75 &   $-$76 \\
 22023$+$5249 &   $-$136    e   &   $-$114    e   &   $-$122    &   $-$139    &   $-$133    &   $-$133    &   $-$131    &   $-$134    &   $-$136    &   $-$133    &   $-$134    \\
 22223$+$4327 &   $-$24 e   &   $-$37 a   &  $-$12 &   $-$27 &   $-$25 &   $-$28 &   $-$28 &   ---  &   ---  &   ---  &   ---  \\
 22272$+$5435 &   $-$28 a   &   $-$38 a   &   $-$35 &   ---     &   $-$32 &   $-$37 &   $-$26 &   ---  &   ---  &   ---  &   ---  \\
 23304$+$6147 &   $-$15 a   &       ---  &    ---    &   $-$16 &   $-$15 &    ---    &    ---    &   ---  &   ---  &   ---  &   ---  \\
\hline
\multicolumn{12}{l}{{\scriptsize P: P-Cygni profile;  a: absorption;
 e: emission; $^*$: different atmospheric and nebular velocities}}\\
\end{tabular}
\end{table*}

\begin{table*}
\caption{Radial velocity measurements (in km s$^{-1}$)
associated to the \ion{Na}{i}D (5889.95 and
5895.92 \AA) and \ion{K}{i} (7698.97 \AA) lines detected in the
post-AGB stars of our sample.}\label{tb:velest2}
\centering
\begin{tabular}{lcccccc}\hline\hline
 & \multicolumn{2}{c}{\ion{K}{i}~7699 \AA} & \multicolumn{2}{c}{\ion{Na}{i} 5890 \AA} & \multicolumn{2}{c}{\ion{Na}{i} 5896 \AA} \\
IRAS Name &   emission  &   absorption  &   emission  &   absorption  &   emission  &   absorption  \\
\hline
 01005$+$7910 $^*$& n.d. &   1,  $-$6  &   20  &   $-$65, $-$41, $-$1    &   21  &   $-$65, $-$42, $-$1    \\
 02229$+$6208   & n.d. &   12,  29 & n.d. &   $-$2,  25 & n.d. &   $-$2,  24 \\
 04296$+$3429 $^*$&   $-$7  &   $-$18,  58    &   21  &   $-$17,  6,  56    &   21  &   $-$16,  6,   56   \\
 05113$+$1347 $^*$& n.d. &   $-$17,  2 &   $-$44 &   $-$17,   6    &   $-$43 &   $-$18,  5 \\
 05251$-$1244   & n.d. &   6   & n.d. &   4,  $-$11 & n.d. &   5,  $-$11 \\
 05341$+$0852   & n.d. &   14  &   23  &   21  &   23  &   21  \\
 06530$-$0213   &    ---   &   --- & n.d. &   25,  47 & n.d. &   25,  46 \\
 07134$+$1005   & n.d. &   93  & n.d. &   29, 39, 91  & n.d. &   28, 40, 92  \\
 08005$-$2356 $^*$& n.d. &   24, $-$8, 48  &   $-$30 &   $-$47,  19    &   $-$30 &   $-$52,  17    \\
 08143$-$4406   & n.d. &   19, 4, 35   & n.d. &   16, 42, 61  & n.d. &   17, 42, 61  \\
 08544$-$4431   & n.d. &   27  & n.d. &   $-$42, 15, 39 & n.d. &   $-$41, 16, 43 \\
 12175$-$5338   & n.d. &   $-$5  &   31  &   $-$27, $-$6, 31 &   32  &   $-$27, $-$5, 32 \\
 16594$-$4656 $^*$& n.d. &   $-$10, $-$41, 5 &   14  &   $-$29, 6  &   13  &   $-$29, 8  \\
 17086$-$2403 $^*$& n.d. &   12  &   $-$6  &   17  &   $-$5  &   14  \\
 17097$-$3210   & n.d. &   5   & n.d. &   $-$29, 3  & n.d. &   $-$29, 3  \\
 17150$-$3224   &   10, 21, 2   & n.d. &   10  &   5   &   10  &   2   \\
 17245$-$3951   &   $-$1,  27 &   $-$86,  $-$57   &   19  &   $-$98, $-$63, $-$31   &   19  &   $-$98, $-$66, $-$31    \\
 17395$-$0841   & n.d. &   17,  3  &   2   &   $-$1  &   3   &   $-$2  \\
 17423$-$1755   & n.d. &   11  &   3, 72   & n.d. &   3, 71   &  n.d.     \\
 17436$+$5003   & n.d. &   $-$50 & n.d. &   $-$66, $-$19    & n.d. &   $-$65, $-$19    \\
 18025$-$3906   & n.d. &   $-$89,  $-$73   & n.d. &   $-$121, $-$34, 8    & n.d. &   $-$121, $-$31, 10   \\
 18062$+$2410   &   ---    &  ---  &   75  &   3, 20   &   67  &   3, 13   \\
 HD 172324      & n.d. &   6   & n.d. &   $-$28, 7  & n.d. &   $-$28, 8  \\
 19114$+$0002   & n.d. &   10, 67, 83  & n.d. &   11, 64, 127 & n.d. &   12, 65, 123 \\
 19200$+$3457   &    ---   &   ---  &   6   &   8   &   7   &   9   \\
 19386$+$0155   & n.d. &   10  &   ---    &  ---  &    ---   &   ---  \\
 19500$-$1709   & n.d. &   4, 22, $-$9   & n.d. &   2, 12, 20   & n.d. &   2, 14, 22   \\
 20000$+$3239   & n.d. &   2,  13  & n.d. &   $-$2,   21    & n.d. &   2,   21 \\
 20462$+$3416   & n.d. &   $-$21,  $-$8    & n.d. &   $-$25, $-$6, 19 & n.d. &   $-$26, $-$7, 17 \\
 22023$+$5249   & n.d. &   $-$23 &   27, $-$130    &   $-$46, 2  &   27, $-$131    &   $-$46, 4  \\
 22223$+$4327   & n.d. &   $-$2, $-$30, $-$43    & n.d. &   $-$42, $-$18, 0  & n.d. &   $-$44, $-$18, 0  \\
 22272$+$5435   & n.d. &   $-$41 & n.d. &   $-$41, $-$4   & n.d. &   $-$41, $-$4   \\
 23304$+$6147 $^*$& n.d. &   $-$1, $-$12, $-$29    &   24  &   $-$52,   $-$29,  $-$7 &   24  &   $-$53,   $-$29,  $-$7 \\
\hline
\multicolumn{7}{l}{{\scriptsize $^*$: emission over absorption; ---:
spectral range not covered; n.d.: non detected}}\\
\end{tabular}
\end{table*}

\begin{table*}
\caption{Radial velocity measurements (in km s$^{-1}$) associated to the DBs
detected in the post-AGB stars of our sample. }
\label{tb:veldib}
\centering
\begin{tabular}{lrrrrrrrrr}\hline\hline
 & \multicolumn{9}{c}{Diffuse band}\\
IRAS Name       &  5780 &   5797        &  5850 &  6196 &  6284 &  6379 &  6614 &  6993 &  7224 \\
\hline
 01005$+$7910   &  $-$16&   29$^\ast$   &   --- & $-$2  &  $-$9 &   2   & $-$12 & $-$16 &   --- \\
 02229$+$6208   &   32  & $-$3$^\ast$   &   --- &   33  &   52  &   22  &   --- &   24  &   --- \\
 04296$+$3429   &   66  &   ---     &   --- &   --- &   55  &   59  &   60  &   --- &   --- \\
 05113$+$1347   &   8   &   ---     &   --- &   --- &   --- &   --- &   0   &   --- &   --- \\
 05251$-$1244   &   $-$3&   11      &   20  &   0   &   29  & $-$14 &   0   &   --- &   --- \\
 05341$+$0852   &   --- &   ---     &   --- &   --- &   --- &   4   &   --- &   15  &   37  \\
 06530$-$0213   &   --- &   ---     &   --- &   --- &   26  &   33  &   12  &   --- &   --- \\
 07134$+$1005   &   $-$5&   $-$14       &   2   &   --- &   --- &   --- &   5   &   --- &   --- \\
 08005$-$2356   &   --- &   30        &   36  &   --- &   --- &   --- &   --- &   --- &   --- \\
 08143$-$4406   &   --- &   ---       &   6   &   --- &   20  &   12  &   24  &   16  &   16  \\
 08544$-$4431   &   23  &   ---       &   16  &$-$45$^\ast$&19&   17  &   28  &   5   &   8   \\
 12175$-$5338   &   11  &   2         &   $-$8& $-$7  &  $-$6 &   --- & $-$4  &   --- &   --- \\
 16594$-$4656   &   --- &   ---       &   9   &   6   &   9   &   6   &   2   & $-$4  &  $-$3 \\
 17086$-$2403   &   14  &   ---       &   --- &   --- &   --- &   --- &   2   &   --- &   --- \\
 17097$-$3210   &   --- &   ---       &   3   &   5   &   5   &   $-$4&   5   &   13  &   0   \\
 17150$-$3224   &   --- &   ---       &   $-$1&   --- &   $-$2&   0   &   9   & $-$3  &   1   \\
 17245$-$3951   &   --- &   ---       &   --- &   --- &   27  &   $-$7&   5   &   16  &   22  \\
 17395$-$0841   &  $-$10&   17        &   --- &   8   &   0   &   3   &   3   &   --- & $-$4  \\
 17423$-$1755   &   15  &   9         &   --- &   9   &   --- &   --- &   12  &   --- &   --- \\
 17436$+$5003   &   --- &   ---     &   23  &   0   &   --- &   19  &   --- &   --- &   --- \\
 18025$-$3906   &   7   & $-$35$^\ast$  &   --- &   --- &   --- &$-$36$^\ast$&24&   19  &   --- \\
 18062$+$2410   &   --- &   15      &   --- &   --- &   --- &   --- &   --- & $-$1  &   --- \\
 HD 172324      &   39  & $-$29$^\ast$  &   --- &   --- &   --- &   --- &   --- &   --- &   --- \\
 19114$+$0002   &   34  &   ---     &   6   &   19  &   12  &   8   &   --- & $-$5  &   29  \\
 19200$+$3457   &   --- &   ---     &   --- &   --- &   --- &   --- &   --- &   --- &   --- \\
 19386$+$0155   &   --- &   ---     &   --- &   5   &   8   &   3   &   6   &   4   &   9   \\
 19500$-$1709   &   11  & $-$44$^\ast$  &   --- &   4   &   6   &   3   &35$^\ast$& --- &   9   \\
 20000$+$3239   &   15  &   ---     &   --- &   6   &   35  &   1   &   --- &   --- &   12  \\
 20462$+$3416   &  $-$14&   $-$4        &45$^\ast$&$-$14&   5   & $-$13 & $-$9  &   --- & $-$2  \\
 22023$+$5249   &  $-$21&   $-$7        &37$^\ast$&$-$14& $-$17 &   2   &   2   &   0   & $-$2  \\
 22223$+$4327   &   17  &   ---     &   --- &   --- &   6   &   2   &   --- &   --- &   --- \\
 22272$+$5435   &   39  &   ---     &   --- &   --- &   --- &   --- &   --- &   --- &   --- \\
 23304$+$6147   &   $-$3&   ---     &   --- & $-$26 &   5   &  $-$8 &   --- &   --- &   --- \\
\hline
\multicolumn{10}{l}{{\scriptsize  $^\ast$ : likely contaminated measurement}}\\
\end{tabular}
\end{table*}

\clearpage

\end{document}